\documentstyle[preprint,aps,graphicx,subfigure]{revtex}
\def\epsp{\epsilon^{'} / \epsilon_K}
\def\eps{\epsilon_K}
\def\lag{\langle}
\def\rag{\rangle}
\def\beq{\begin{equation}}
\def\eeq{\end{equation}}
\def\barr#1{\begin{array}{#1}}
\def\earr{\end{array}}
\def\beqar{\begin{eqnarray}}
\def\eeqar{\end{eqnarray}}
\def\beqars{\begin{eqnarray*}}
\def\eeqars{\end{eqnarray*}}
\def\bitem{\begin{itemize}}
\def\eitem{\end{itemize}}

\def\bc{\begin{center}}
\def\ec{\end{center}}
\def\beq{\begin{equation}}
\def\eeq{\end{equation}}
\def\bea{\begin{eqnarray}}
\def\eea{\end{eqnarray}}
\def\bit{\begin{itemize}}
\def\eit{\end{itemize}}
\def\ben{\begin{enumerate}}
\def\een{\end{enumerate}}
\def\ba{\begin{array}}
\def\ea{\end{array}}
\def\bc{\begin{center}}
\def\ec{\end{center}}

\def\ol{\overline}

\def\wt{\widetilde}
\def\al{\alpha}
\def\be{\beta}

\def\ga{\gamma}

\def\eps{\varepsilon}

\def\dReg(#1,#2){\SetOffset(#1,#2)\BCirc(0,0){4}
    \Line(2.8,2.8)(-2.8,-2.8)
    \Line(2.8,-2.8)(-2.8,2.8)\SetOffset(0,0)}

\newenvironment{rowvec2}{\left(\begin{array}{cc}}{\end{array}\right)}
\newenvironment{colvec}{\left(\begin{array}{c}}{\end{array}\right)}
\def\bcol{\begin{colvec}}
\def\ecol{\end{colvec}}
\def\brow2{\begin{rowvec2}}
\def\erow2{\end{rowvec2}}

\newcommand{\mg}{m_{\tilde{g}}}
\newcommand{\lesssimilar}
{\mbox{\raisebox{-.5ex}{$\stackrel{\mbox{$<$}}{\sim}$}}}
\newcommand{\mtext}[1]{\rm{#1}}
\begin{document}
\draft
\preprint{\begin{tabular}{l}
\hbox to\hsize{May, 2001 \hfill KAIST-TH 2001/06}\\[-3mm]
\hbox to\hsize{ \hfill  }\\[5mm] \end{tabular} }

\bigskip

\title{Gluino-squark contributions to \\ 
CP violations in the kaon system }
\author{$^a$ S. Baek, $^b$ J.-H. Jang, $^c$ P. Ko and $^c$ J.~H.~Park}
\address{$^a$ Dep. of Physics, National Taiwan University,  
Taipei 10764, Taiwan
\\
$^b$ Institute of Photonics, Electronics, and Information Technology,\\
Chonbuk National University, Chonju Chonbuk 561-756, Korea
\\
$^c$ Dep. of Physics, KAIST, Taejon 305-701, Korea}
\maketitle
\tighten
\begin{abstract}
Recently it was shown, within the mass insertion approximation (MIA), that 
the gluino mediated flavor changing neutral current (FCNC) interactions can 
saturate both $\epsilon_K$ and ${\rm Re} ( \epsilon' / \epsilon_K )$ 
even for the real Cabibbo--Kobayashi--Maskawa (CKM) matrix 
through a single CP violating parameter 
$( \delta_{12}^d )_{LL}$. 
In this work, we extend our previous analysis to the nonvanishing KM phase, 
and to the effective SUSY model where the MIA is no longer a good 
approximation.
In our model, $\epsilon_K$ and $B^0 - \overline{B^0}$ mixing can receive 
significant SUSY contributions. Therefore the usual constraints on $\rho$ and 
$\eta$ in the Wolfenstein parametrization of the CKM matrix can be relaxed
except for the $| V_{ub}/ V_{cb}|$ from the semileptonic $b\rightarrow u$
transition. 
This affects 
$K \rightarrow \pi \nu\bar{\nu}$ and $K_L \rightarrow \mu^+ \mu^-$
indirectly, even if they are not directly induced by gluino-mediated FCNC,
while $K_L \rightarrow \pi^0 e^+ e^-$ is influenced both directly and
indirectly.
Differences between 
our model and other SUSY models enhancing 
${\rm Re} ( \epsilon' / \epsilon_K )$ are discussed in detail.
\end{abstract}


\newpage
\narrowtext
 \tighten


\section{Introduction}
\label{sec:intro}

For many years, CP violating phenomena was observed only in 
$K_{L} \rightarrow 2 \pi$ \cite{cronin}, which could be attributed to 
$K^0 - \overline{K^0}$ mixing  ($\Delta S = 2$). The CP violating parameter 
in the $K^0 - \overline{K^0}$ mixing, $\epsilon_K$, has been accurately 
measured :
$\epsilon_K = e^{i \pi/4} ~( 2.271 \pm 0.017 ) \times 10^{-3}$ \cite{pdg}. 
However, the recent observation of ${\rm Re} (\epsilon' / \epsilon_K )$ by 
KTeV  collaboration, 
$
{\rm Re} ( \epsilon' / \epsilon_K ) = ( 28.0 \pm 4.1 ) \times 10^{-4}
$
\cite{ktev}, 
nicely confirms the earlier NA31 experiment \cite{na31}
$
{\rm Re} ( \epsilon' / \epsilon_K ) = (23 \pm 7) \times 10^{-4}.
$
Including another new data from  NA48 collaboration, $( 14.0 \pm 4.3 ) \times 
10^{-4}$ \cite{na48}, the current world average for 
${\rm Re} (\epsilon' / \epsilon_K )$  is
\begin{equation}
{\rm Re} (\epsilon' / \epsilon_K ) = ( 19.2 \pm 2.4 ) \times 10^{-4}.
\end{equation}
This nonvanishing number indicates unambiguously the existence of  CP 
violation in the decay amplitude $(\Delta S = 1)$, and the original form of 
superweak model is excluded. Along with the measurements of $\sin 2 \beta$ 
from time-dependent asymmetry in $B^0 \rightarrow J/\psi K_S$ at $B$ 
factories \cite{b2psik}, we are now in the blooming era for detailed studies 
of CP violations in $K$ and $B$ systems. However, in this article, we will 
consider CP violations in the $K$ system only for the reason to be explained 
later.

Two parameters $\epsilon_K$ and ${\rm Re} ( \epsilon' / \epsilon_K ) $
that quantify CP violations in the kaon system can be accommodated by the 
Kobayashi--Maskawa (KM) phase 
in the Glashow--Salam--Weinberg's standard model (SM). However, the 
SM predictions for the latter is rather uncertain because of large 
uncertainties in the nonperturbative matrix elements and the strange quark 
mass. There are significant variations in theoretical predictions 
\cite{buras1,bertolini,paschos,pich,truong}, 
which indicate the current status of our understanding of 
${\rm Re} (\epsilon' / \epsilon_K )$.  In this work, we will follow closely 
the work by Bosch {\it et al.} \cite{buras1} in calculating  
${\rm Re} (\epsilon' / \epsilon_K )$ and other rare kaon decays so that we 
can compare the numerical results with other results more easily.  

Although the SM can accommodate a bulk of the observed ${\rm Re} ( \epsilon' 
/ \epsilon_K )$ within certain theoretical uncertainties, it is also
interesting to speculate possible new physics contributions to CP violations 
in the kaon system. Among various scenarios beyond the SM, 
the minimal supersymmetric standard model (MSSM) and its various extensions 
are most promising candidates. Thus it is quite natural to consider generic 
SUSY contributions to $\epsilon_K$ and ${\rm Re} ( \epsilon' / \epsilon_K )$.
In this regard, two of us (SB and PK) showed that the flavor preserving 
$A_t$ and $\mu$ phases in the effective SUSY models alone cannot generate 
large enough $\epsilon_K$ \cite{abel,baek1,masiero} 
or new phase shift in 
$B^0 - \overline{B^0}$ mixing beyond that coming from the KM phase 
\cite{baek2}. For $\epsilon^{'} / \epsilon_K$, there have been a lot of works
done after the KTeV announcement \cite{keum,he,chanowitz,mm,babu1,babu2,khalil,ko1,brhlik0,strumia,buras99,nir,kn,isidori99}. 
Those works in the supersymmetric models can be divided into three categories :
\begin{itemize}
\item the enhanced $sdg$ vertex 
\cite{mm,babu1,babu2,khalil,ko1,brhlik0,strumia,buras99,nir}
\item the enhanced $sdZ$ vertex \cite{buras99,isidori99}
\item the enhanced EW penguins \cite{kn}.
\end{itemize}
Models in the first category are divided further into two parts, depending on
the origin of the flavor and CP violations : either from the flavor changing
squark masses in the $LL$ sector or the flavor changing $LR$ sector due to
nontrivial flavor structures in the trilinear $A$ terms which may arise 
naturally in string inspired models with $D$ branes. 
Our model presented in Ref.~\cite{ko1} is unique compared to other models
with enhanced $sdg$ vertex, since SUSY CP violations in our model affects 
not only the ${\rm Re} ( \epsilon' / \epsilon_K )$ but also the $\epsilon_K$ 
by significant amounts, contrary to other models in this category.  
In fact, it was shown that both $\epsilon_K$ and 
${\rm Re} ( \epsilon' / \epsilon_K )$ can be saturated by a single complex 
number $(\delta_{12}^d )_{LL}$, if $| \mu \tan\beta |$ is large enough $\sim 
{\cal O}(10)$ TeV so that the induced $(\delta_{12}^d )_{LR}$ through 
the double mass insertion can be ${\cal O}(10^{-5})$. Furthermore, 
there is no conflict with the neutron electric dipole moment (EDM) 
in our model.
There is another work showing that all the CP violations observed so far can
be accommodated by a single CP violating phase in the gluino mass parameter
$M_{\tilde{g}}$ \cite{brhlik0}. They assumed a very specific flavor structure 
in the trilinear $A$ couplings in order to generate large 
$\epsilon^{'} $. On the other hand, we assume a specific flavor structure
in the left squark mass matrix only, and the flavor structure and CP 
violating phase of a trilinear $A$ coupling is totally irrelevant in our case.

In generic SUSY models, there can be potentially large contributions to 
flavor changing neutral currents (FCNC)
and CP violating processes from gluino-squark mediations, which is 
termed as SUSY flavor/CP problems \cite{kolda}.  Since these effects are
due to strong interactions, it may be parametrically large compared to the
charged Higgs and chargino contributions, unless the squarks and gluinos are 
too heavy. There are basically three ways to solve SUSY flavor problems : 
(i) universality of squark masses, (ii) alignment of squark 
and quark mass matrices due to some flavor symmetries, and (iii) decoupling
of the 1st/2nd generation squarks. 

In order to study the gluino mediated flavor changing phenomena within the 
frameworks of approximate universality or alignment, it is convenient to 
use the so-called mass insertion approximation (MIA) \cite{mia}. 
In this approximation, one works in the super CKM basis where the quark mass 
matrices are diagonal.
The quark-squark-gluino vertex is flavor diagonal in the MIA, and
the flavor/chirality mixing occur through the insertion of 
$( \delta_{ij}^d )_{AB}$, where $i,j=1,2,3$ and $A,B=L,R$ denote the flavors 
of the squarks under consideration and the chiralities of their superpartners. 
The superscript denotes that the down type squark mass matrix is involved.
The parameters $( \delta_{ij}^d )_{AB}$ characterize the size of the 
gluino-mediated flavor changing amplitudes, and they may be CP violating  
complex numbers, in general. In the following, diagrams involving charged 
Higgs, chargino and neutralino will be ignored, since they are suppressed by 
$\alpha_w /\alpha_s$ compare to the gluino-squark loops unless gluino/squarks 
are very heavy. This should be a good starting point for studying the SUSY 
FCNC/CP problems. 

In addition to the MIA, we also consider the vertex mixing (VM) model 
which is a
good approximation if the third generation sfermions are lighter than the 
1st/2nd generation sfermions as in the effective SUSY models \cite{decouple}.  
In this case, the SUSY FCNC and CP problems are solved by the decoupling 
(and some degeneracy in the 1st/2nd generation sfermion masses),  and we  
keep only the left/right sbottoms and gluino contributions to the 
$\Delta S=1,2$ effective Hamiltonians.  The related works in $B$ physics can 
be found in Refs.~\cite{kkl}.

Although SUSY CP violations can saturate both $\epsilon_K$ and 
${\rm Re} (\epsilon' / \epsilon_K )$, it would be unnatural to assume that  
$\delta_{KM}$ vanishes (or equals $\pi$), since there is no symmetry 
principle supporting such assumption \footnote{We can impose CP symmetry to be
broken only softly. This case is already included in our model with 
$\delta_{KM} = 0$ or $\pi$. See Refs.~\cite{dine} for more discussions.}. 
In this work, we present more detailed analysis of our model within both MIA 
and VM cases with nonvanishing $\delta_{KM}$.  Since the $\epsilon_K$ and 
$B^0 - \overline{B^0}$ mixing are affected by SUSY contributions 
significantly in our model, the usual 
constraints on $( \rho, \eta )$ from these two quantities need not be held 
any more. Only the constraint from $b\rightarrow u$ semileptonic decay would 
be valid. Therefore, when we vary the $( \rho, \eta )$, we will impose this 
constraint only, and the apex of the unitarity triangle (UT) can be 
anywhere in 
the $( \rho, \eta )$ plane, not only in the first quadrangle. This is in 
sharp contrast to other analysis within SUSY models where new physics 
contributes significantly only to ${\rm Re} (\epsilon' / \epsilon_K )$ 
or the analysis based on the MSSM with minimal flavor violation.  
There is an indication that the unitarity triangle does not collapse 
(namely $\alpha, \beta, \gamma$ do not vanish) just from determining the 
triangle only from the unitarity of 
the Cabibbo--Kobayashi--Maskawa (CKM) matrix \cite {ckm} :
\begin{equation}
\alpha = 19^{\circ} - 142^{\circ}, ~~~
\beta  = 6^{\circ} - 31^{\circ}, ~~~
\gamma = 28^{\circ} - 152^{\circ}. 
\end{equation}
Also there are some indications that charmless nonleptonic 2--body decays 
of $B$ mesons seem to prefer $\cos\gamma <0$ \cite{hou,burasf}, although 
theoretical uncertainties involved are of vastly different degrees. 
Considering all these points, we consider all the possibility, although 
$\gamma \sim \pi$ may be unrealistic.
 
Assuming that the $\epsilon_K$ and $B^0 - \overline{B^0}$ mixing cannot be 
used to constrain the CKM elements, we study several rare $K$ decays such as 
$K\rightarrow \pi \nu\bar{\nu}$, $K\rightarrow \pi l^+ l^-$ and 
$K \rightarrow \mu^+ \mu^-$ including the squark--gluino contributions, and 
compared with other recent works on ${\rm Re} (\epsp)$. 
We do not consider the CP asymmetries in $K \rightarrow 3 \pi$,
$K \rightarrow \pi\pi \gamma $ and hyperon decays. 
These observables were discussed in the literatures 
within the context of the enhanced $sdg$ vertex as the origin of large 
$\epsp$. In our model, we have another source of FCNC and CP violation, namely
$( \delta_{12}^d )_{LL}$ which induces $\Delta S = 1$ four-quark operators.
The detailed study of these effects to  $K \rightarrow 3 \pi$,
$K \rightarrow \pi\pi \gamma $ and hyperon decays 
are beyond the scope of the present work.  Also these processes will suffer 
from large hadronic uncertainties as other nonleptonic kaon decays in 
chiral perturbation theory, unlike $K\rightarrow \pi \nu\bar{\nu}$. 

The contents of this paper are organized as follows.
In the Sec.~II A and B, we review briefly the $\Delta S =2$ and $\Delta S =1$ 
effective Hamiltonians and their relations to $\epsilon_K$ and $\epsp$.
In Sec.~II C and D, we recapitulate the basic formulae for neutron EDM, 
and branching ratios and CP asymmetries for 
$K\rightarrow \pi \nu\bar{\nu}$, $K\rightarrow \pi l^+ l^-$ and 
$K \rightarrow \mu^+ \mu^-$.
The numerical inputs for various parameters in our model are summarized 
in Sec.~II E. In Sec.~III and IV, we discuss the CP violation phenomenology 
in the kaon sector for the MIA and the VM cases, respectively. Finally we 
summarize in Sec.~V.

\section{Effective Hamiltonians for kaon physics}
\label{sec:frame}

In order to discuss $\epsilon_K$, $\epsilon'/\epsilon_K$ and other rare 
kaon decays within the SM and SUSY models, we first construct the effective 
Hamiltonians for $\Delta S= 2$ and $\Delta S = 1 $. 
We obtain the Wilson coefficients of the effective theory after matching
it to the full theory at ${\cal O}(m_W)$. In this section, we define the 
operator basis for these effective Hamiltonians and the recipes for the 
renormalization group (RG) running from the weak scale to the lower energy
scale $\mu \approx m_c$.  The relevant Wilson coefficents for the 
$\Delta S = 1$ and $\Delta S= 2$ effective Hamiltonians are
explicitly given in the Secs.~III and IV, where we describe the MI and 
the VM approximations for the CP and flavor-violating effects of SUSY in
detail.

\subsection{Effective Hamiltonian for $\Delta S = 2$ and $\epsilon_K$}
\label{subsec:eps}

The effective Hamiltonian describing the $\Delta S =2$ $K^0 - \overline{K^0}$
mixing in the SM and the MSSM (with only the gluino contributions included) 
can be written as 

\begin{equation}
     H_{\rm eff}^{\Delta S = 2}= H_{\rm SM}^{\Delta S = 2} +
                             H_{\rm SUSY}^{\Delta S = 2}
\end{equation}
where \cite{buras95,buras98}
\begin{eqnarray}
   H_{\rm SM}^{\Delta S = 2} & = & \frac{G_F^2}{16 \pi^2} m_W^2 \left[
   \lambda_c^{*2} \eta_1^{(SM)} S_0(x_c) + \lambda_t^{*2} 
   \eta_2^{(SM)} S_0(x_t)+
   2 \lambda_c^* \lambda_t^* \eta_3^{(SM)} S_0(x_c,x_t) \right]
\nonumber 
\\
 & \times &    \left[ \alpha_s (\mu) \right]^{-2/9} ~
 \bar{d}^{\alpha} \gamma_{\mu} (1-\gamma_5) s^{\alpha}
 \bar{d}^{\beta} \gamma^{\mu} (1-\gamma_5) s^{\beta}
 ~+~ \rm h.c. ,
\\
H_{\rm SUSY}^{\Delta S = 2} & = &
\frac{1}{216 \: \wt{m}^2}
\left(\sum_{i=1}^{5} C_i^{(2)} Q_i + \sum_{i=1}^{3} \wt{C}_i^{(2)} \wt{Q}_i
\right) ,
\end{eqnarray}
with operators defined to be
   \beqar
   Q_1&=& \alpha_s^2 (\mu) \:
         \bar{d}^{\alpha} \gamma_{\mu} (1-\gamma_5) s^{\alpha}
         \bar{d}^{\beta} \gamma^{\mu} (1-\gamma_5) s^{\beta},
         \nonumber \\
   Q_2&=& \alpha_s^2 (\mu) \:
         \bar{d}^{\alpha} (1-\gamma_5) s^{\alpha}
         \bar{d}^{\beta} (1-\gamma_5) s^{\beta},
         \nonumber \\
   Q_3&=& \alpha_s^2 (\mu) \:
         \bar{d}^{\alpha} (1-\gamma_5) s^{\beta}
         \bar{d}^{\beta} (1-\gamma_5) s^{\alpha},
         \\
   Q_4&=& \alpha_s^2 (\mu) \:
         \bar{d}^{\alpha} (1-\gamma_5) s^{\alpha}
         \bar{d}^{\beta} (1+\gamma_5) s^{\beta},
         \nonumber \\
   Q_5&=& \alpha_s^2 (\mu) \:
         \bar{d}^{\alpha} (1-\gamma_5) s^{\beta}
         \bar{d}^{\beta} (1+\gamma_5) s^{\alpha}.
         \nonumber 
   \eeqar         
Here, $\wt{m}$ is the average squark mass in MIA, or the sbottom mass in VM\@.
$\alpha$ and $\beta$ are color indices.  The operators $\tilde{Q}_i$ are 
obtained from $Q_i$ by interchanging $L$ and $R$. 
The explicit form of the loop function, 
$S_0(x,y)$, can be found in Ref. \cite{buras95}.

We have separated the effective Hamiltonian into two parts :
one coming from the SM and the other from SUSY sector, because the former 
is of ${\cal O}( \alpha_w )$,  
whereas the latter is of ${\cal O}( \alpha_s )$.
We separately do RG running of SUSY and SM contributions to get
the Wilson coefficients for the kaon physics at ${\cal O} (m_c)$.  
In the SM part, we use the well-known anomalous dimesion matrices
for RG running which can be found in Ref.\cite{buras95}. 
For the RG running of SUSY part, we use a modified method advocated in  
\cite{Borzumati:1999qt} in which new effective operators are defined
including  the $\alpha_s^n$ factor coming from the gluino-squark 
loop diagrams. 

After defining the new operator set, we can apply the RG 
procedure of SM to gluino mediated contributions: {\it matching} 
calculation for Wilson coefficents to order $\alpha_s^0$ and 
{\it RG evolution} by using the anomalous dimesion matrix to order 
$\alpha_s^1$ for leading log approximations (LLA).
The calculation of matrix elements is obviously modified by the
additional factors of new operators.

The anomalous dimension matrix in our basis at leading order is given by
\begin{equation}
\label{ano_dim_ds1}
\gamma^{(0)} = \widetilde{\gamma}^{(0)} + 4~\beta_0^{(f)}~\mbox{\bf 1},
\end{equation}
where $\beta_0^{(f)} = (11 - 2 f /3)$ with the effective flavor number of 
$f$ and $\widetilde{\gamma}^{(0)}$ can be obtained in \cite{bagger97},
\begin{equation}
   \widetilde{\gamma}^{(0)} = \left( \begin{array}{rrrrr}
    4 &     0 &    0 &   0 &  0 \\
    0 & -28/3 &  4/3 &   0 &  0 \\
    0 &  16/3 & 23/3 &   0 &  0 \\
    0 &     0 &    0 & -16 &  0 \\
    0 &     0 &    0 &  -6 &  0  \end{array} \right) .
\end{equation}
The Wilson coefficients at ${\cal O}(m_c)$ running down from those
at ${\cal O}(m_W)$ read
\begin{eqnarray}
   C_1^{(2)} (m_c) &=& \eta_1 C_1^{(2)} (m_W) , \nonumber \\
   C_2^{(2)} (m_c) &=& \eta_{22} C_2^{(2)} (m_W) 
+ \eta_{23} C_3^{(2)} (m_W) , \nonumber \\
   C_3^{(2)} (m_c) &=& \eta_{32} C_2^{(2)} (m_W) 
+ \eta_{33} C_3^{(2)} (m_W) , \\
   C_4^{(2)} (m_c) &=& \eta_4 C_4^{(2)} (m_W) 
+ \frac{1}{3} \left( \eta_4 -  
                 \eta_5 \right) C_5^{(2)} (m_W) , \nonumber \\
   C_5^{(2)} (m_c) &=& \eta_5 C_5^{(2)} (m_W) , \nonumber 
\end{eqnarray}
with
\begin{eqnarray}
   \eta_1 &=&  \left( \frac{\alpha_s (m_b)}{\alpha_s (m_c)} \right)^{56/25}
               \left( \frac{\alpha_s (m_W)}{\alpha_s (m_b)} \right)^{52/23} ,
               \nonumber \\
   \eta_2 &=&  \left( \frac{\alpha_s (m_b)}{\alpha_s (m_c)} \right)^{1.419}
               \left( \frac{\alpha_s (m_W)}{\alpha_s (m_b)} \right)^{1.369} ,
               \nonumber \\
   \eta_3 &=&  \left( \frac{\alpha_s (m_b)}{\alpha_s (m_c)} \right)^{2.661}
               \left( \frac{\alpha_s (m_W)}{\alpha_s (m_b)} \right)^{2.718} ,
               \\
   \eta_4 &=&  \left( \frac{\alpha_s (m_b)}{\alpha_s (m_c)} \right)^{26/25}
               \left( \frac{\alpha_s (m_W)}{\alpha_s (m_b)} \right)^{22/23} ,
               \nonumber \\
   \eta_5 &=&  \left( \frac{\alpha_s (m_b)}{\alpha_s (m_c)} \right)^{53/25}
               \left( \frac{\alpha_s (m_W)}{\alpha_s (m_b)} \right)^{49/23} ,
               \nonumber 
\end{eqnarray}
and
\begin{equation}
   \begin{array}{lll}
   \eta_{22} =  0.983 \eta_2 + 0.017 \eta_3 ,&~~~& 
   \eta_{23} = -0.258 \eta_2 + 0.258 \eta_3 , \\
   \eta_{32} = -0.064 \eta_2 + 0.064 \eta_3 ,&~~~& 
   \eta_{33} =  0.017 \eta_2 + 0.983 \eta_3 .
   \end{array}
\end{equation}
   


In order to estimate $\Delta M_K$ and $\epsilon_K$, it is necessary to know
the hadronic matrix elements: $ \lag \bar{K}^0 | Q_i (\mu) | K^0 \rag $
for $\mu \sim m_c$. In the present work, we use the following expressions
\cite{gabbiani96,ciuchini98},
which are sufficient for our purpose in the LLA. 


   \beqar             
   \lag K^0 | Q_1(\mu) | \ol{K^0} \rag &=& 
   \frac{8}{3} \: \alpha_s^2 (\mu) \:m_K^2\:f_K^2\:B_1(\mu),
   \nonumber \\
   \lag K^0 | Q_2(\mu) | \ol{K^0} \rag &=& - \frac{5}{3} \: \alpha_s^2 (\mu)
   \left(\frac{m_K}{m_s(\mu)+m_d(\mu)}\right)^2 m_K^2\:f_K^2\:B_2(\mu),
   \nonumber \\
   \lag K^0 | Q_3(\mu) | \ol{K^0} \rag &=& \frac{1}{3} \: \alpha_s^2 (\mu)
   \left(\frac{m_K}{m_s(\mu)+m_d(\mu)}\right)^2 m_K^2\:f_K^2\:B_3(\mu),
   \nonumber \\
   \lag K^0 | Q_4(\mu) | \ol{K^0} \rag &=& \alpha_s^2 (\mu)
   \left[ \frac{1}{3} + 2
   \left(\frac{m_K}{m_s(\mu)+m_d(\mu)}\right)^2\right] 
   m_K^2 \: f_K^2\:B_4(\mu),
   \nonumber \\
   \lag K^0 | Q_5(\mu) | \ol{K^0} \rag &=& \alpha_s^2 (\mu)
   \left[ 1 + \frac{2}{3} 
   \left(\frac{m_K}{m_s(\mu)+m_d(\mu)}\right)^2\right] 
   m_K^2 \: f_K^2\:B_5(\mu).
   \eeqar         
The bag parameters $B$'s are given by \cite{ciuchini98}
\beqar             
   B_1(\mu)&=& 0.60, \nonumber \\
   B_2(\mu)&=& 0.66, \nonumber \\
   B_3(\mu)&=& 1.05, \nonumber \\
   B_4(\mu)&=& 1.03, \nonumber \\
   B_5(\mu)&=& 0.73, 
\eeqar         
at $\mu=2$ GeV ( We use these values at $\mu=m_c=1.3$GeV).

The CP violation in the $K^0 - \overline{K^0}$ mixing is characterized by 
a complex parameter $\epsilon_K$, which can be obtained from the 
$\Delta S =2$ effective Hamiltonian by the following formula:
\beq
\epsilon_K=\frac{\mbox{exp}(i\pi/4)}{\sqrt{2}\Delta M_K}
 \left[ {\rm Im} M_{12} + 2 \xi {\rm Re} M_{12} \right] ,
\eeq
where $M_{12}$ is defined as 
\beq
2 m_K M_{12} = \lag K^0 | H_{{\rm eff}}^{\Delta S=2} | \ol{K^0} \rag .
\eeq
Remember that $\xi={\rm Im} A_0/{\rm Re} A_0$ is very small and thus
is neglected in our calculation.


\subsection{Effective Hamiltonian for $\Delta S = 1$ and Re $(\epsp)$}
\label{subsec:epsp}

The effective Hamiltonian describing the $\Delta S =1$ kaon decays
receive contributions from the SM and 
reads (with only the gluino contributions included):
   \beq             
   H_{\rm eff}^{\Delta S = 1} = 
   H_{\rm SM}^{\Delta S = 1} + H_{\rm SUSY}^{\Delta S = 1} ,
   \eeq
   where
   \begin{eqnarray}
   H_{\rm SM}^{\Delta S = 1} &=&
   C'_1 O'_1 + C'_2 O'_2 +
   \sum_{i=3}^{10}  C'_i O'_i  +
   \mbox{h.c.} , \\
   \label{ds1_susy}
   H_{\rm SUSY}^{\Delta S = 1} &=& C_1 O_1 + C_2 O_2 +
   \sum_{i=3}^{10} ( C_i O_i + \wt{C}_i \wt{O}_i ) +
   \sum_{A=\pm} \sum_{B=\gamma,g} \sum_{X=s,\tilde{g}} C^A_{B X} O^A_{B X} +
   \mbox{h.c.}
   \end{eqnarray}
   The explicit form of $O'_i$ can be refered in Ref. \cite{buras95} and
   new operators $O_i$'s are defined as 
   \begin{equation}
   \label{new_ds1}
   O_i (\mu) = \alpha_s^2 (\mu)~O^{'}_i(\mu).
   \end{equation}
   $\tilde{O}_i$ are obtained from $O_i$ by interchanging
   $1-\gamma_5$ and $1+\gamma_5$. The $10 \times 10$ anomalous dimension
   matrix for RG evolution in these new four-quark operators is defined in 
   a  similar way  to that of $\Delta S = 2$ case given
   in Eq. (\ref{ano_dim_ds1}).
   The last part of Eq.(\ref{ds1_susy}) is for the magnetic and chromomagnetic
   operators in which the subindex $B$ denotes whether the chirality flips
   on the strange quark or the gluino line:
   \beqar
   \label{magnetic_op}
   O_{\gamma s}^{\pm} 
   &=& m_s \frac{e \: \alpha_s (\mu)}{32 \pi^2} \left[ \bar{d} 
    \sigma^{\mu\nu} F_{\mu\nu} (1+\gamma_5) s \pm
    \bar{d} \sigma^{\mu\nu} F_{\mu\nu} (1-\gamma_5) s \right],
   \nonumber \\
   O_{\gamma \tilde{g}}^{\pm} 
   &=& \phantom{m_s} \frac{e \: \alpha_s (\mu)}{32 \pi^2} \left[ \bar{d} 
    \sigma^{\mu\nu} F_{\mu\nu} (1+\gamma_5) s \pm
    \bar{d} \sigma^{\mu\nu} F_{\mu\nu} (1-\gamma_5) s \right],
   \nonumber \\
   O_{g s}^{\pm} 
   &=& m_s \frac{g \: \alpha_s (\mu)}{32 \pi^2} \left[ \bar{d} 
    \sigma^{\mu\nu} t^a G_{\mu\nu}^a (1+\gamma_5) s \pm
    \bar{d} \sigma^{\mu\nu} t^a G_{\mu\nu}^a (1-\gamma_5) s \right],
    \nonumber \\
   O_{g \tilde{g}}^{\pm} 
   &=& \phantom{m_s} \frac{g \: \alpha_s (\mu)}{32 \pi^2} \left[ \bar{d} 
    \sigma^{\mu\nu} t^a G_{\mu\nu}^a (1+\gamma_5) s \pm
    \bar{d} \sigma^{\mu\nu} t^a G_{\mu\nu}^a (1-\gamma_5) s \right].
    \nonumber
   \eeqar         
   
The anomalous dimension matices for RG evolution of the above equations are
given by \cite{Borzumati:1999qt}
\begin{eqnarray}
  \gamma^{(0)}_{\tilde{g}} & = &
  \widetilde{\gamma}^{(0)} + (2 \beta^{(f)}_0 - 8) \: {\bf 1} ,
  \\
  \gamma^{(0)}_s & = &
  \widetilde{\gamma}^{(0)} + 2 \beta^{(f)}_0 \: {\bf 1} , \\
  \widetilde{\gamma}^{(0)} & = &
  \left(
  \begin{array}{rr}
     32/3 & 0 \\
    -32/9 & 28/3
  \end{array}
  \right) .
\end{eqnarray}

 The Wilson coefficients of magnetic operators at ${\cal O} (m_c)$ are
 \beqar
 C_{\gamma (\tilde{g},s)} (m_c) &=& \eta^{(\tilde{g},s)}_1~
      C_{\gamma (\tilde{g},s)} (m_W) - \frac{8}{3} 
      \left( \eta^{(\tilde{g},s)}_1 - \eta^{(\tilde{g},s)}_2 \right)
      C_{g (\tilde{g},s)} (m_W) , \nonumber \\ 
 C_{g (\tilde{g},s)} (m_c) &=& \eta^{(\tilde{g},s)}_2~C_{g (\tilde{g},s)} (m_W) ,
 \eeqar
 with
 \begin{equation}
 \eta^{\tilde{g}}_1 = \left( \frac{\alpha_s (m_b)}{\alpha_s (m_c)} 
      \right)^{29/25} \left( \frac{\alpha_s (m_W)}{\alpha_s (m_b)} 
      \right)^{27/23}, ~~~~
 \eta^{\tilde{g}}_2 = \left( \frac{\alpha_s (m_b)}{\alpha_s (m_c)} 
      \right)^{27/25} \left( \frac{\alpha_s (m_W)}{\alpha_s (m_b)} 
      \right)^{25/23},
 \end{equation}
 and
 \begin{equation}
 \eta^s_1 = \left( \frac{\alpha_s (m_b)}{\alpha_s (m_c)} 
      \right)^{41/25} \left( \frac{\alpha_s (m_W)}{\alpha_s (m_b)} 
      \right)^{39/23}, ~~~~
 \eta^s_2 = \left( \frac{\alpha_s (m_b)}{\alpha_s (m_c)} 
      \right)^{39/25} \left( \frac{\alpha_s (m_W)}{\alpha_s (m_b)} 
      \right)^{37/23}. 
 \end{equation}
     
  The hadronic matrix element of a four quark operator in 
  Eq. (\ref{new_ds1})
  is obtained by mutiplying the corresponding one of SM 
  in Ref.~\cite{buras95} by $\alpha_s^2(m_c)$. For the magnetic operators, the
  matrix elements at ${\cal O}(m_c)$ can be expressed as \cite{buras99}
  \beqar
  \langle O_{g s}^- \rangle_0 =
   m_s \langle O_{g \tilde{g}}^- \rangle_0 
   &=& m_s \: \alpha_s (\mu) \:
   \sqrt{\frac{3}{2}} \frac{11}{16 \pi^2} 
   \frac{\langle \overline{q}q \rangle}{F_\pi^3 } m_\pi^2 B_G,
   ~~~~~ \lag O_g^+ \rag_0 = 0, 
  \eeqar
  where the subscript 0 means $\Delta I = 0$ components of the matrix elements.

The parameter ${\rm Re} (\epsilon'/\epsilon_K )$, which characterizes the 
direct CP violation in the decay amplitude of $K \rightarrow \pi\pi$, is 
expressed as following in terms of the $\Delta S = 1$ effective Hamiltonian:
\beq
{\rm Re} \left(\frac{\epsilon'}{\epsilon_K}\right)=
-
\frac{\omega}{\sqrt{2} |\epsilon| {\rm Re} A_0} 
\sum_{i} {\rm Im}(C_i) \left( \lag O_i \rag_0 ( 1- \Omega_{\eta+\eta'})
- \frac{1}{\omega} \lag O_i \rag_2 \right)
\eeq
where
$\omega = {\rm Re} A_2 / {\rm Re} A_0 = 0.045$, and 
$\Omega_{\eta+\eta'} = 0.25 \pm 0.05$ 
represents the isospin breaking effects.

\subsection{Effective Hamiltonian for the neutron EDM}
\label{subsec:edm}

The electric dipole moment of a fermion $f$
can be described by the effective Hamiltonian \cite{brhlik}
\begin{equation}
  H_{\rm eff}^{\rm EDM} = \sum_{i = 1}^3 \: C_i^{\rm edm} \: O_i^{\rm edm} ,
\end{equation}
and its basis operators
\begin{eqnarray}
  O_1^{\rm edm} & = & - \frac{i}{2} \bar{f} \sigma^{\mu\nu} \ga_5 f F_{\mu\nu} , 
  \nonumber \\
  O_2^{\rm edm} & = & - \frac{i}{2} \bar{f} \sigma^{\mu\nu} \ga_5 T^a f G^a_{\mu\nu} , 
  \nonumber \\
  O_3^{\rm edm} & = & - \frac{1}{6} f_{abc}
  G^a_{\mu\rho} G^{b\rho}_\nu G^c_{\lambda\sigma} 
  \epsilon^{\mu\nu\lambda\sigma} .
\end{eqnarray}
With this Hamiltonian, we calculate the (chromo)EDM of quarks and gluons  
and compare the result to the experimental value to constrain possible 
parameter space.
The current experimental upper bound on the neutron EDM is 
\cite{Harris:1999jx}
\begin{equation}
  \label{eq:neutron-edm-bound}
  | d_n | \: \lesssimilar \: 6.3 \times 10^{-26} \: e \: {\rm cm} .
\end{equation}


\subsection{Rare kaon decays : $K\rightarrow \pi\nu\bar{\nu}, 
~ \pi l^+ l^-, ~\mu^+ \mu^- $}

%
%

SUSY contributions to various rare kaon decays through the (chromo)magnetic
and enhanced $sdZ$ operators have been discussed comprehensively by Buras
{\it et al.} \cite{buras99}, and we heavily lean on formulae for branching 
ratios for rare kaon decays given in their paper. In the numerical analysis,
however, there are two major different points to be emphasized : 
\begin{itemize}
\item In our case, we ignore the enhanced $sdZ$ operator, which may be 
important  only in a restricted parameter space, as already noticed in 
Ref.~\cite{buras99}. Therefore, our model is precisely the Scenario A therein.

\item When we vary CKM matrix elements $\lambda_t$, we impose constraints 
coming from semileptonic decays only, since these are fairly insensitive to
new physics beyond the SM, especially on the SUSY contributions we are 
considering here. On the other hand, $\epsilon_K$ and $B^0_{d(s)} - 
\overline{B^0_{d(s)}}$ mixing may be affected by new physics. Indeed gluino 
mediated FCNC present in our model can affect both of these observables
in a significant manner, unlike those considered in most recent works related
with $\epsilon^{'} / \epsilon_K$ in the framework of SUSY. Since we consider
the kaon sector only and the $B$ meson sector is independent of the kaon 
sector in the MIA, we do not impose the $B^0 - \overline{B^0}$ mixing
in this work. Therefore, the weak phase $\gamma = \phi_3$ can be anywhere 
between 0 and $2 \pi$. 
\end{itemize}

Also $( \delta_{12}^d )_{LL}$ would contribute to $s\rightarrow d q\bar{q}$
penguin operators, thereby modifying nonleptonic kaon decays (and CP 
violations therein) such as $K \rightarrow 3 \pi$ and 
$K \rightarrow \pi\pi \gamma$ and hyperon decays. The effects of  
$( \delta_{12}^d )_{LR}$ on these decays have been considered in the 
literatures, but not those of $( \delta_{12}^d )_{LL}$. Since these will 
involve some hadronic uncertainties compared to rare kaon decays we consider
here (such as $K \rightarrow \pi\nu\bar{\nu}$), we will not include these
nonleptonic kaon and hyperon decays in this work.

In the SM, $K \rightarrow \pi \nu \bar{\nu}$ is affected by the KM phase.
In fact the neutral mode $K_L \rightarrow \pi^{\circ} \nu \bar{\nu}$ is 
purely CP violating so that its branching ratio would vanish in the limit
of exact CP symmetry. In our model with two sources of CP violations, only
$\delta_{KM}$ affects these decays, since $s \rightarrow d g$ vertex does 
not contribute to them. 

Using the formulae given in Ref.~\cite{buras99}
we have the results
\begin{eqnarray}
  B( K_L \rightarrow \pi^0 \nu \bar{\nu} ) & = &
  B^0_{\rm SM} =
  6.78 \times 10^{-4} ( X_0 {\rm Im} \lambda_t )^2 , 
  \label{eq:bkpnn0}
  \\
  B( K^+ \rightarrow \pi^+ \nu \bar{\nu} ) & = &
  B^+_{\rm SM} =
  1.55 \times 10^{-4} \left[
    ( X_0 {\rm Im} \lambda_t )^2 +
    ( X_0 {\rm Re} \lambda_t + \Delta_c )^2
    \right] ,
    \label{eq:bkpnn+}
\end{eqnarray}
where \( X_0 = C_0 - 4 B_0 = 1.52 \)
is a combination of loop functions evaluated at 
\( \bar{m}_t (m_t) = 166~{\rm GeV} \),
and \( \Delta_c =  - ( 2.11 \pm 0.30 ) \times 10^{-4} \) 
is the internal charm contribution.


Unlike $K\rightarrow \pi \nu\bar{\nu}$, the direct CP violating part of 
$K_L \rightarrow \pi^0 e^+ e^-$ is affected in our model, 
because the Wilson coefficients of $O_{\gamma (\tilde{g}, s)}$ and 
$O_{7V} \equiv ( \bar{d} s)_{(V-A)} (\bar{e} e)_{V}$ 
are modified due to the new CP violating phase from
the squark--gluino diagrams.
As for other SUSY contributions, we work with the `SUSY version'
of operator
\begin{equation}
O^{\rm SUSY \prime}_{7V} 
\equiv \alpha \: \alpha_s(\mu) \: ( \bar{d} s)_{(V-A)} (\bar{e} e)_{V} ,
\end{equation}
rather than $O_{7V}$, 
in order to do matching and running formally consistent with LLA.
Under the definition,
\begin{equation}
  {\rm Im} \Lambda_g^+ \tilde{y}_\gamma =
  - \frac{\alpha_s B_T}{\sqrt{2} G_F m_K}
    \left[
    m_s \: {\rm Im} C_{\gamma s}^+ +
    {\rm Im} C_{\gamma \tilde{g}}^+
    \right] ,
\end{equation}
the branching ratio is given by
\begin{eqnarray}
  \label{eq:kpee}
  B( K_L \rightarrow \pi^0 e^+ e^- )_{\mtext{dir}}
  & = & 6.3 \times 10^{-6} \times \nonumber \\
  & & \Bigg[
    \left( 
      {\rm Im} \lambda_t \tilde{y}_{7V} + 
      {\rm Im} \Lambda_g^+ \tilde{y}_\gamma +
      \frac{\sqrt{2}}{G_F} 2\pi \alpha_s(m_c) 
      {\rm Im} C^{\rm SUSY \prime}_{7V}
    \right)^2 \\
    & & \phantom{\Bigg[} + ({\rm Im} \lambda_t \tilde{y}_{7A})^2 
  \Bigg] , \nonumber
\end{eqnarray}
where, for \( \Lambda_{\overline{MS}} = 325 \: {\rm MeV} \) and
\( m_t = 170 \: {\rm GeV} \), \cite{buras95}
\begin{eqnarray}
  \tilde{y}_{7V} ( \mu = 1.3 \: {\rm GeV} ) & = & \phantom{-} 0.537 \times 2\pi, \nonumber\\
  \tilde{y}_{7A} ( \mu = 1.3 \: {\rm GeV} ) & = & - 0.700 \times 2\pi . \nonumber
\end{eqnarray}


The short distance part of the branching ratio of
\( K_L \rightarrow \mu^+ \mu^- \) within our model
is the same as that in the SM \cite{buras99}:
\begin{equation}
  B( K_L \rightarrow \mu^+ \mu^- )_{\rm SD} =
  B^{\mu \mu}_{\rm SM} =
  6.32 \times 10^{-3} ( Y_0 {\rm Re} \lambda_t + \ol{\Delta}_c )^2 ,
\end{equation}
where
\(
  \ol{\Delta}_c = - ( 6.54 \pm 0.60 ) \times 10^{-5}
\)
is the charm contribution.

\subsection{Input parameters}
\label{subsec:para}

Since there may be large SUSY contributions to $\epsilon_K$ as well as 
to $B^0 - \overline{B^0}$ in our model, one cannot use the constraints on
the CKM matrix elements coming from these observables. We impose only the 
constraint coming from the semilpetonic decays of mesons. 
Therefore we have \cite{pdg}
\begin{equation}
| V_{ub} / V_{cb} | = 0.08 \pm 0.02
\end{equation}
This leaves the phase of $V_{ub}$ undetermined so that the SM contributions
to various observables are not fully determined.
In the following numerical analysis, we sample data fixing the phase
of $V_{ub}$ to some selected values of $0^{\circ}, 60^{\circ}, 120^{\circ}...$
so on up to $360^\circ$.
%
%

We list numerical input values used throughout our analysis 
in Table~\ref{tab:input}.

%




%




\section{Mass Insertion Approximation Case}
\label{sec:mia}

\subsection{Approximate flavor symmetry and $( \delta_{12}^d )_{AB}$}

One way to solve the gluino-mediated FCNC problems in general SUSY models
is to assume that the squark masses are approximately diagonal and degenerate 
in the basis where quark mass matrices are diagonal (the so-called super CKM
basis). 
\[
( \tilde{m}^2_{ij} )_{AB} = ( \tilde{m}^2 )_{AB} \delta_{ij}
+ ( \delta m^2_{ij} )_{AB}.  
\]
In this case, one can expand the squark propagator around the common
squark mass matrix $( \tilde{m}^2 )_{AB} \delta_{ij}$, and the rest 
can be cast into interaction Hamiltonians. As long as $\delta m^2$ is small
compared to the common squark mass $\tilde{m}^2$, one can keep the lowest 
order perturbations in $\delta m^2$ relevant to the processes we are 
interested in. In many cases, just a single mass insertion along a single 
squark propagator is sufficient, but double mass insertion along a single 
squark propagator should be included in some cases. In fact, it is the case
for $\epsilon^{'} / \epsilon_K$ in the relatively large $\mu \tan\beta$ 
region \cite{ko1}. In this case, smallness of $s$ quark mass can be 
compensated 
by a large $\mu \tan\beta$ (which has been ignored in most previous 
literatures) so that an appreciable amount of $\tilde{s}_R - \tilde{s}_L$ may 
be possible. 

The typical size of $(\delta_{ij}^d)_{AB} \equiv ( \delta m^2_{ij} 
)_{AB} / \tilde{m}^2$ should be fairly small in order to be consistent with
$\Delta M_K$ and $\epsilon_K$. Theoretical understanding of such a small 
number constitutes the so--called SUSY flavor problems. This problem could be
solved by approximate universality or approximate flavor symmetry where both 
quarks and squarks are almost aligned in flavor space.

The gluino--squark contributions to the $\Delta S=1,2$ effective Hamiltonian
at the heavy SUSY particle scale have been calculated by several groups,
and are encoded in the Wilson coefficients listed below. 

\subsection{Wilson coefficient for $\Delta S =2$}

  The Wilson coefficients for the 
  \( \Delta S = 2 \) effective
  Hamiltonian at the SUSY particle mass scales 
  coming from the gluino contributions are given as follows:
   \beqar             
   C^{(2)}_1&=&-\frac{1}{4} \left( 24 x f_6(x)+ 66 \wt{f}_6(x) \right)
         \left( \delta_{12}^d \right)^2_{LL}, \nonumber \\
   C^{(2)}_2&=&C^{(2)}_3=C^{(2)}_4=C^{(2)}_5=\wt{C}^{(2)}_1=0, \nonumber \\
   \wt{C}^{(2)}_2&=&-\frac{816}{4} x f_8(x) 
   \left( \delta_{12}^d \right)^2_{LL} \left( \delta_{22}^d \right)^2_{LR},
   \nonumber \\
   \wt{C}^{(2)}_3&=&\frac{144}{4} x f_8(x) \left( \delta_{12}^d \right)^2_{LL} 
   \left( \delta_{22}^d \right)^2_{LR}.
   \eeqar         
The expressions for $\wt{C}^{(2)}_2  $ and $\wt{C}^{(2)}_3 $ are new, 
but their 
contributions to $\epsilon_K$ and $\Delta m_K$ are negligible.  
The loop functions are given by the following expressions :
   \beqar             
   f_6(x)&=& \frac{6(1+3x)\ln x+x^3-9x^2-9x+17}{6(x-1)^5},
   \nonumber \\
   \wt{f}_6(x)&=&\frac{6x(1+x)\ln x-x^3-9x^2+9x+1}{3(x-1)^5},
   \\
   f_8(x) &=& \frac{197+25 x-300 x^2 + 100 x^3 - 25 x^4 + 3 x^5+60(1+5x) \ln x}
        {60 (x-1)^7}, \nonumber
   \eeqar
where $ x = ( m_{\tilde{g}}/\tilde{m} )^2$.  
The flavor conserving but chirality changing mass insertion parameter 
$( \delta_{22}^d )_{LR}$ is defined as
   \beqar
   ( \delta_{22}^d )_{LR} &=& \frac{m_s}{\wt{m}^2} (A_s^* -\mu \tan
   \beta)
   \eeqar         
   We neglect the terms
   proportional to $( \delta_{12}^d )^2_{LR}$,
   $( \delta_{12}^d )^2_{RL}$, and
   $( \delta_{12}^d )^2_{RR}$.

\subsection{Wilson coefficient for $\Delta S =1$}

The Wilson coefficients for the $\Delta S =1$ effective Hamiltonian at the 
SUSY particle mass scales are given as follows :
   \beqar             
   C_i &=& 0~~~~~~ \mbox{(for $i$ = 1,2,7--10)}\nonumber \\
   C_3&=& \frac{1}{4 \wt{m}^2}
   \left(-\frac{1}{9} B_1(x)-\frac{5}{9} B_2(x)-\frac{1}{18} P_1(x)
   -\frac{1}{2} P_2(x) \right) 
   \left(\delta^d_{12}\right)_{LL} \nonumber \\
   C_4&=& \frac{1}{4 \wt{m}^2}
   \left(-\frac{7}{3} B_1(x)+\frac{1}{3} B_2(x)+\frac{1}{6} P_1(x)
   +\frac{3}{2} P_2(x) \right) 
   \left(\delta^d_{12}\right)_{LL} \nonumber \\
   C_5&=& \frac{1}{4 \wt{m}^2}
   \left(\frac{10}{9} B_1(x)+\frac{1}{18} B_2(x)-\frac{1}{18} P_1(x)
   -\frac{1}{2} P_2(x) \right) 
   \left(\delta^d_{12}\right)_{LL} \nonumber \\
   C_6&=& \frac{1}{4 \wt{m}^2}
   \left(-\frac{2}{3} B_1(x)+\frac{7}{6} B_2(x)+\frac{1}{6} P_1(x)
   +\frac{3}{2} P_2(x) \right) \left(\delta^d_{12}\right)_{LL} 
    \nonumber \\
   \wt{C}_i&=& C_i ~\mbox{with replacement of $LL$ by $RR$},
   \quad i = 3, 4, 5, 6
   \eeqar         
   where 
   \beqar             
   B_1(x)&=&\frac{1+4x-5x^2+4x\mbox{log} x +2x^2\mbox{log} x}{8(1-x)^4}
   \nonumber \\
   B_2(x)&=&x\frac{5-4x-x^2+2\mbox{log} x +4x\mbox{log} x}{2(1-x)^4}
\nonumber \\
   P_1(x)&=&\frac{1-6x+18x^2-10x^3-3x^4+12x^3\mbox{log}x}{18(x-1)^5}
   \nonumber \\
   P_2(x)&=&\frac{7-18x+9x^2+2x^3+3\mbox{log}x-9x^2\mbox{log}x}{9(x-1)^5}
   \eeqar         
   $B_i$ and $P_i$ are box and penguin diagram  contributions respectively.


   \beqar
   C_{\gamma s}^\pm 
   &=& - \frac{8 \pi Q_d}{3~ \wt{m}^2}
   \left(\delta_{12}^d \right)_{LL} M_4(x) , \\
   C_{\gamma \tilde{g}}^\pm 
   & = &
   \phantom{-} \frac{8 \pi Q_d}{3~ \wt{m}^2}
   \left(\delta_{12}^d \right)_{LL} \left(\delta_{22}^d \right)_{LR}
    \left( \wt{m} \sqrt{x} \right) M_2(x) ,
   \\
   C_{g s}^\pm 
   &=& -\frac{2 \pi}{\wt{m}^2} \left(\delta_{12}^d
   \right)_{LL} \left(\frac{3}{2} M_3(x) -\frac{1}{6} M_4(x) \right) ,
   \\
   C_{g \tilde{g}}^\pm 
   & = &
   \phantom{-} \frac{2 \pi}{\wt{m}^2}
   \left(\delta_{12}^d \right)_{LL} \left(\delta_{22}^d \right)_{LR}
   \left( \wt{m} \sqrt{x} \right)
   \left( \frac{3}{2} M_1(x) -\frac{1}{6} M_2(x) \right)
   \eeqar
   where $Q_d=-1/3$ and
   \beqar
   M_1(x)&=& \frac{3-3x^2+(1+4 x+ x^2)\ln x}{(x-1)^5}
   \nonumber \\
   M_2(x)&=& \frac{1+9x-9x^2-x^3+6 x (1+x)\ln x}{2(x-1)^5}
   \\
   M_3(x)&=& \frac{1}{3} M_2(x)
   \nonumber \\
   M_4(x)&=& \frac{-1+9x+9x^2-17x^3+6x^2 (3+x)\ln x}{12(x-1)^5}
   \nonumber 
   \eeqar
Here again, the Wilson coefficients $ C_{\gamma \tilde{g}}^\pm $ and
$ C_{g \tilde{g}}^\pm $ arising from double mass insertions are given
for the first time. These two new terms distinguish our model from other 
models, when we discuss the processes $K_L \rightarrow \pi^0 e^+ e^-$ and
$K_L \rightarrow \mu^+ \mu^-$ through $s\rightarrow d \gamma$, and the 
process ${\rm Re}~(\epsilon^{'}/\epsilon_K)$ through $s\rightarrow d g$.
The process $K_L \rightarrow \pi^0 e^+ e^-$ is also affected by
the $s \rightarrow d l^+ l^-$ local interaction whose Wilson
coefficient at the SUSY particle mass scale is
\begin{equation}
  C^{\rm SUSY \prime}_{7V} =
  - \frac{4}{9 \tilde{m}^2} P_1 (x) (\delta^d_{12})_{LL}.
\end{equation}
    
\subsection{EDM constraint}

The Wilson coefficients for the effective Hamiltonian for
the neutron EDM in MIA are given by
\begin{eqnarray}
  C_1^{\rm edm} & = & - \frac{2}{3} \: \frac{e \al_s}{\pi} \: Q_d \:
  \frac{\mg}{\wt{m}^2} \: {\rm Im} \left(\delta_{11}^d \right)_{LR} \:
  4 \: B_1 (x) , \\
  C_2^{\rm edm} & = & - \frac{1}{4} \: \frac{g_s \al_s}{\pi} \:
  \frac{\mg}{\wt{m}^2} \: {\rm Im} \left(\delta_{11}^d \right)_{LR} \:
  \left(\frac{3}{x} B_2 (x) - \frac{4}{3} B_1 (x) \right) , 
\end{eqnarray}
where
   \begin{equation}
     ( \delta_{11}^d )_{LR} = \frac{m_d}{\wt{m}^2} (A_d^* -\mu \tan \be) .
   \end{equation}

   If we consider the loop functions to be order of one,
   and set both the squark and the gluino masses to \( 500 \: {\rm GeV} \),
   we get the limit on the flavor preserving mixing parameter,
   \begin{equation}
     \label{eq:Imd11LR}
     {\rm Im} (\delta_{11}^d)_{LR} \: \lesssimilar \:
     {\cal O} (10^{-8}) ,
   \end{equation}
   from the constraint~(\ref{eq:neutron-edm-bound}).
   The typical modulus of \( (\delta_{11}^d)_{LR} \)
   is ${\cal O} (10^{-4})$
   for \( A_d^* - \mu \tan \be = 10 \: {\rm TeV} \),
   and this limit constrains the phase of \( (\delta_{11}^d)_{LR} \)
   below ${\cal O} (10^{-4})$.

   There is a correlation between phases of
   \( (\delta_{11}^d)_{LR} \)
   and
   \( (\delta_{22}^d)_{LR} \)
   because they share a common term $ \mu \tan \be $,
   and this correlation becomes stronger if
   we assume universality between  $A_d$ and $A_s$.
   This indirectly constrains the phase of 
   \( (\delta_{22}^d)_{LR} \).
   But we can safely set its phase to zero while satisfying
   $\epsilon_K$ and ${\rm Re} (\epsilon'/\epsilon_K )$ at the same time,
   as we will show in the numerical analysis.



\subsection{Numerical Results}

\begin{figure}
  \begin{center}
      \includegraphics[width=17cm]{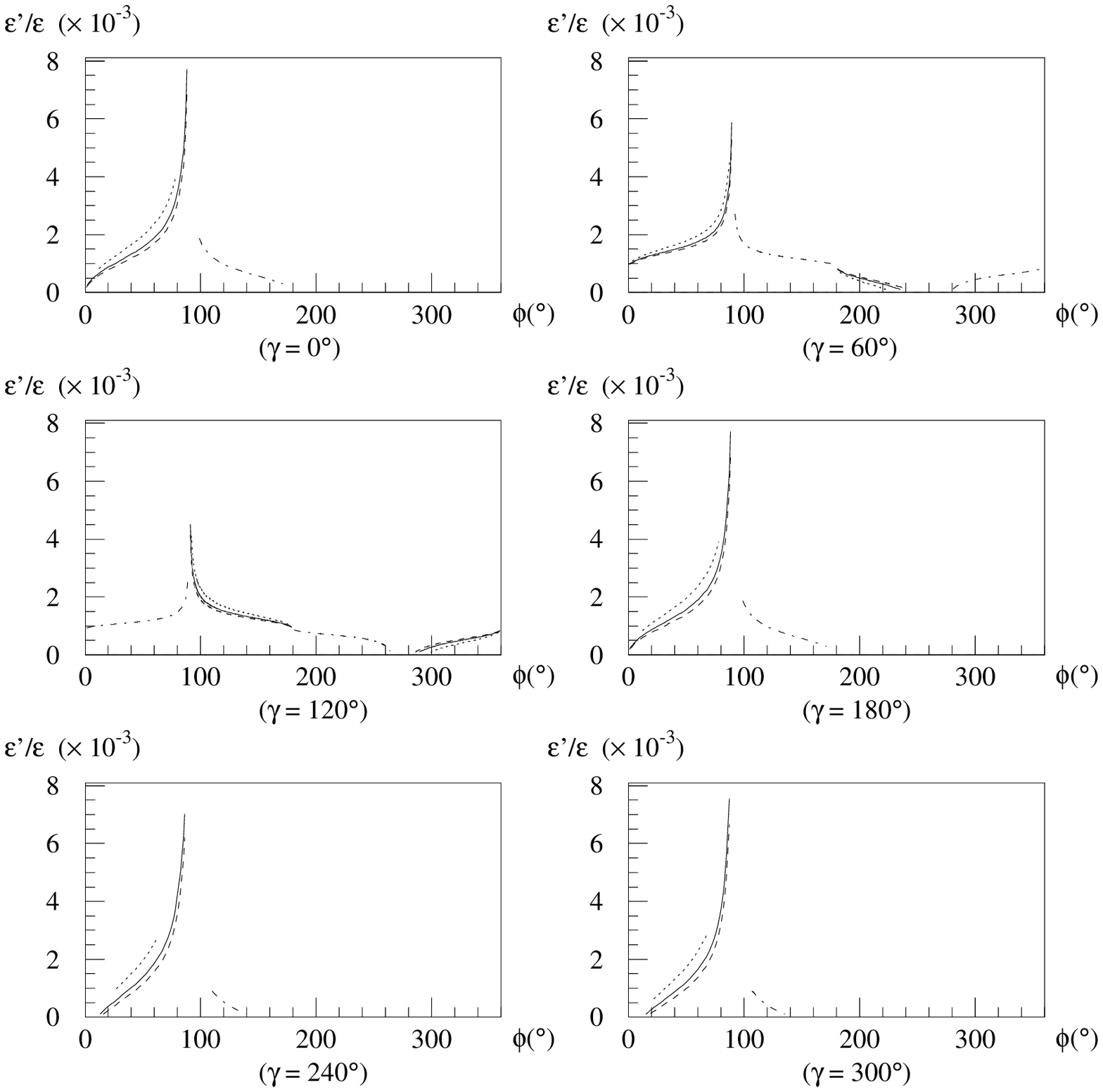}
    \caption{
      The plots of
      $ \epsilon' / \epsilon $ versus the phase of
      $ (\delta_{12}^d)_{LL} $ in MI for
      six different values of $\ga$ with
      $A_s^* - \mu \tan \be = 10 \: {\rm TeV} $.
      Graphs were drawn in the solid, the dashed, the dotted,
      and the dash-dotted lines
      for $ x = 0.3, 1.0, 2.0, 4.0 $, respectively.
      }
    \label{fig:epsp-phase-mi5}
  \end{center}
\end{figure}

 Now in Fig.~\ref{fig:epsp-phase-mi5}, we show the plots of 
\( {\rm Re}~( \epsilon' / \epsilon_K ) \) as a
function of the phase of \( (\delta_{12}^d)_{LL} \), 
where \( (\delta_{12}^d)_{LL} \) is parametrized as \( r e^{i \phi} \).
We fixed $A_s^* - \mu \tan\beta = 10$ TeV. Different values and sign of
$A_s^* - \mu \tan\beta $ yield similar results. For a given value of
$\phi$, $r$ is determined in such a way that $\epsilon_K$ becomes the
experimental value, and \( {\rm Re} ( \epsilon' / \epsilon_K) \) is computed
from those two parameters. As $\phi$ varies from $180^\circ$ to $360^\circ$
for $\gamma = 0^{\circ}$, the SUSY contribution to \( {\rm Re} 
( \epsilon' / \epsilon_K ) \) repeats its values in the region of $\phi$ from
$0^\circ$ to $180^\circ$,  with the sign flipped. Therefore it is clearly 
ruled out, and the results are not shown. But as the angle $\ga$ varies, 
the SM contribution either increases or decreases 
\( {\rm Re} ( \epsilon' / \epsilon_K ) \), so that the invisible parts in 
$\ga = 0$ graph float up in $\ga = 60^\circ$ and $\ga = 120^\circ$ graphs, 
while some visible parts sink down in $\ga = 240^\circ$ and 
$\ga = 300^\circ$ graphs.
Note that the functional relation between $r$ and $\phi$ governed by the  
$\epsilon_K$ constraint depends on the KM angle $\ga$, and thus plots for 
nonzero $\ga$ are not simple shifts of that for vanishing $\ga$.

Since there may be large contributions to $K^0 - \overline{K^0}$ and
$B - \overline{B^0}$ mixing in our model, it is utterly important to 
measure three angles $\alpha, \beta$ and $\gamma$ of the unitarity triangle 
(UT) in a way indepedent of these neutral meson mixings, in addition to the 
more popular ways using  $B\rightarrow J/\psi K_S, \pi \pi$, so on.
Also, it would be an interesting and important question how much portions of 
\( {\rm Re} ( \epsilon' / \epsilon_K ) \) come from the SM and new physics 
(SUSY gluino--squark loops in this work). This question can be answered only 
if the angle $\ga$ is determined by some other methods independent of 
$B^0 - \overline{B^0}$ mixing, e.g., from $B\rightarrow D K$ or 
$B\rightarrow \pi K$, etc.. Once the angle $\ga$ is known, one can calculate 
the SM contribution, \( {\rm Re} ( \epsilon' / \epsilon_K )^{\rm SM} \). 
Then, the difference between the observed 
\( {\rm Re} ( \epsilon' / \epsilon_K ) \)  and the SM would be the new 
physics contributions.  



\begin{figure}
  \begin{center}
    \subfigure[]%
    {\includegraphics[width=7.5cm]{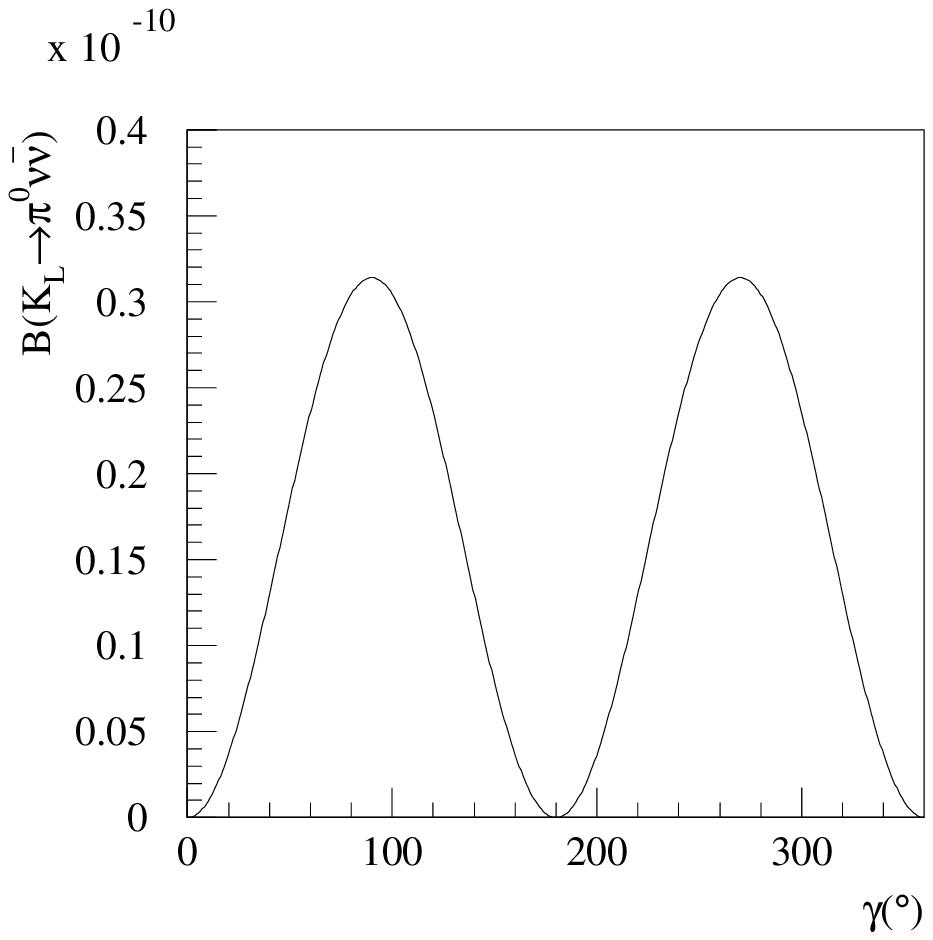}}
    \subfigure[]%
    {\includegraphics[width=7.5cm]{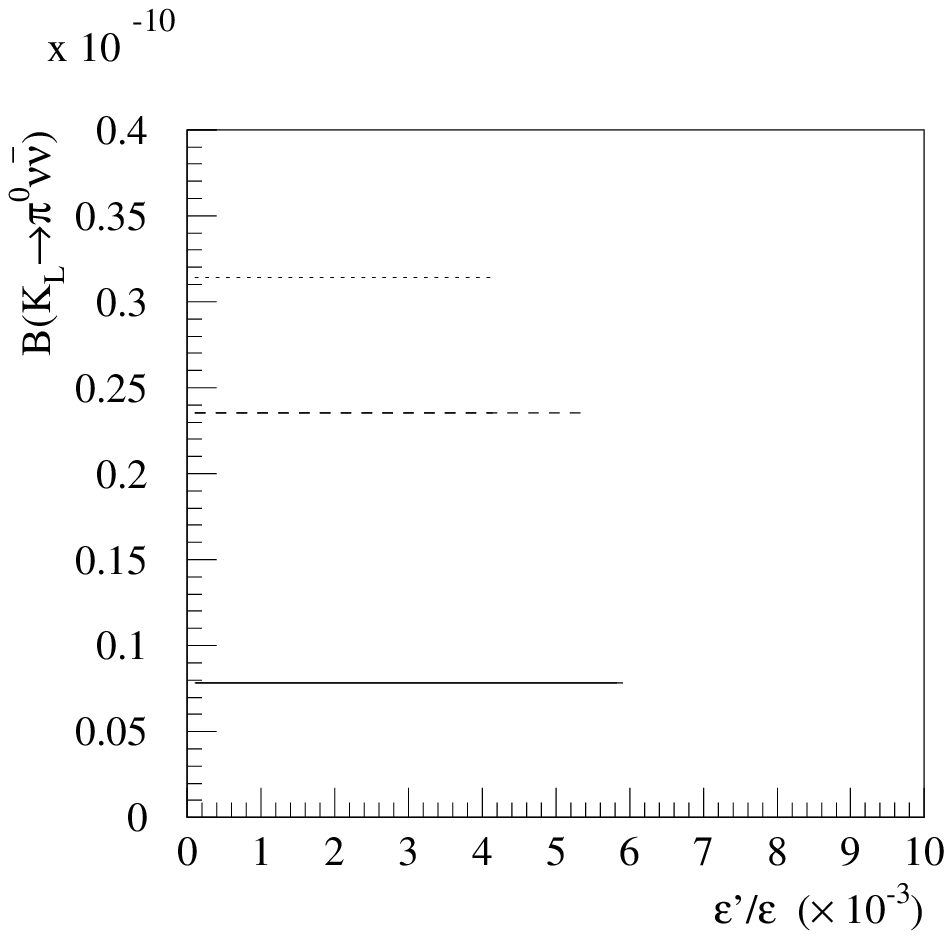}} \\
    \subfigure[]%
    {\includegraphics[width=7.5cm]{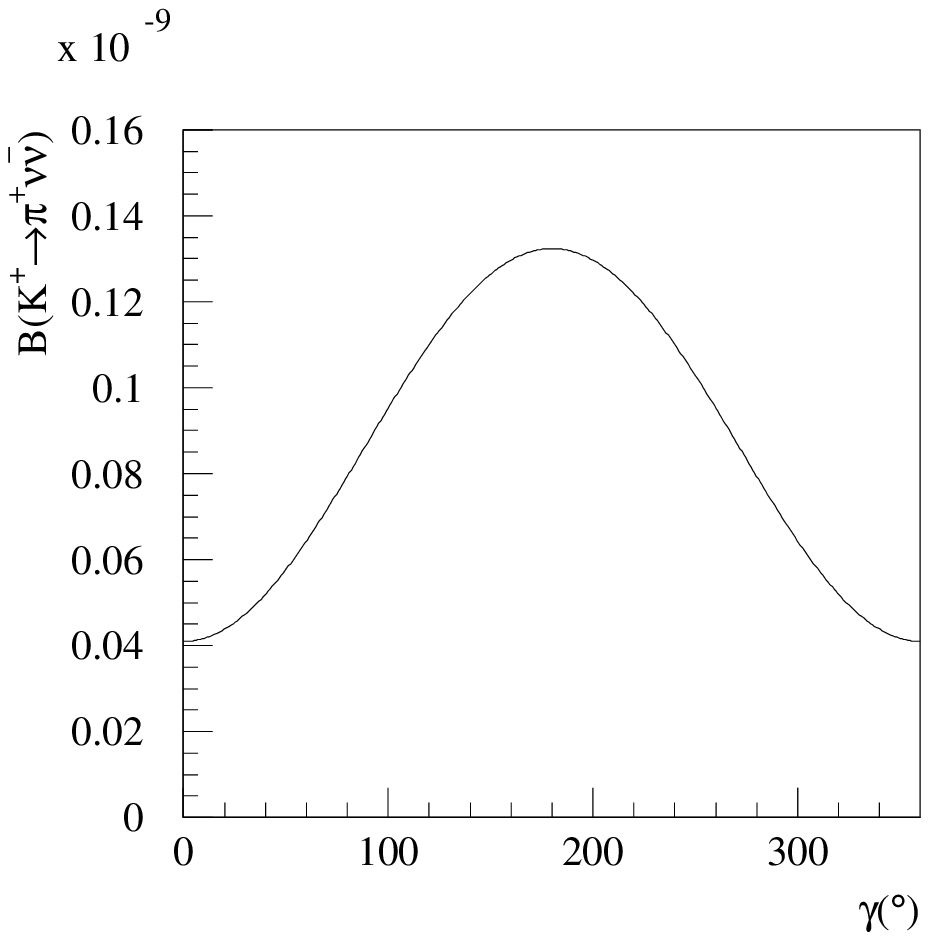}}
    \subfigure[]%
    {\includegraphics[width=7.5cm]{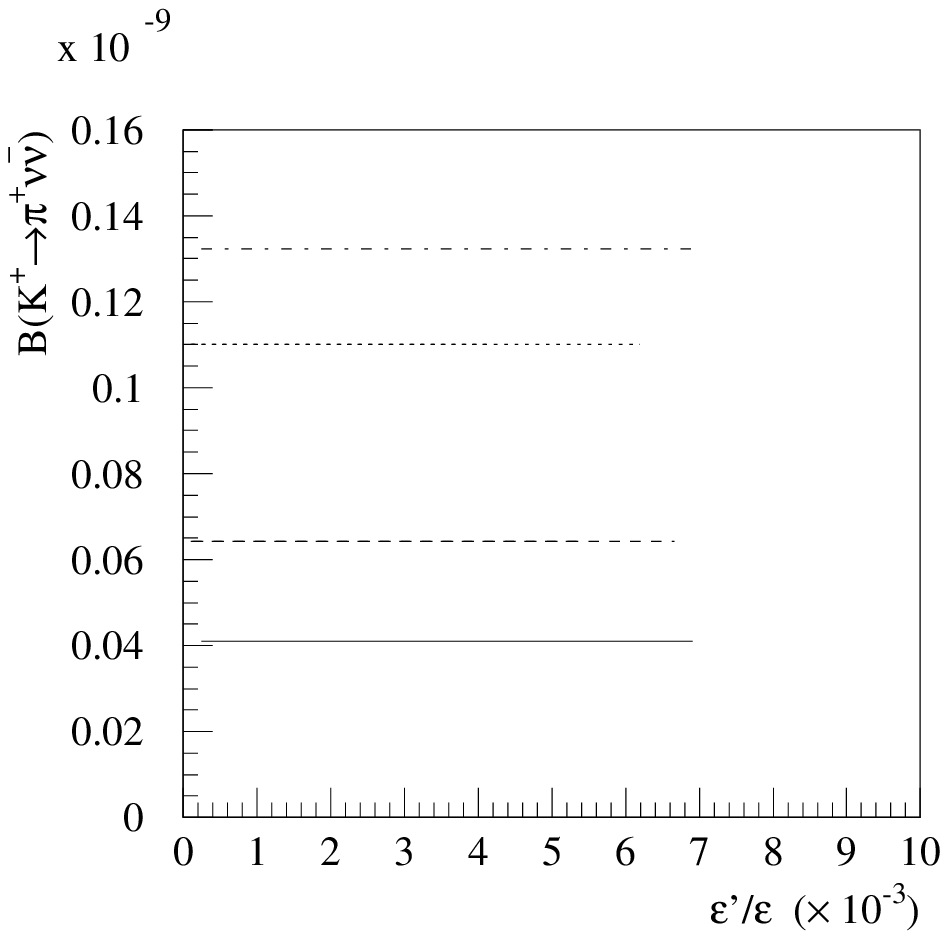}}
    \caption{
      The $\gamma$ dependences of $ B( K \rightarrow \pi \nu \bar{\nu} ) $
      ((a), (c))
      and their correlations with $ \epsilon' / \epsilon $ in MI ((b), (d)),
      for $x = 1$ with
      $A_s^* - \mu \tan \be = 10 \: {\rm TeV} $.
      (b) The solid, the dashed, and the dotted lines correspond to
      $ \ga = 30^\circ, 60^\circ, 90^\circ $, respectively.
      The first two equally correspond to $ \ga = 150^\circ, 120^\circ $
      as well.
      The branching ratio vanishes for
      $ \ga = 0^\circ \mbox{\@ or \@} 180^\circ $.
      (d) The solid, the dashed, the dotted, and the dash-dotted lines
      correspond to
      $ \ga = 0^\circ, 60^\circ, 120^\circ, 180^\circ $,
      respectively, and $360^\circ$ minus them.
      }
    \label{fig:kpnn-mi5}
  \end{center}
\end{figure}

In our model, $K \rightarrow \pi \nu \bar{\nu}$ decays are not directly
affected by the gluino--squark loops. This is because neutrinos couple to 
neither of photons and gluinos, and thus the gluino--squark loop do not 
induce $sd\nu\bar{\nu}$ vertex. Still, the shape of the UT, namely the angle 
$\gamma$, can be different from the SM, because of potentially large 
contributions to \( \epsilon_K \) (and also to $B^0 -\overline{B^0}$ mixing). 
Therefore, gluino mediated FCNC will affect the branching ratios for 
\( K \rightarrow \pi \nu \bar{\nu} \)  in an indirect way, even if the Wilson 
coefficients for $s\rightarrow d \nu \bar{\nu}$ remains the same as in the SM.
In Fig.~\ref{fig:kpnn-mi5} (a) and (b), we show the
$K_L \rightarrow \pi^0 \nu \bar{\nu}$ branching ratio as a function of 
$\gamma$ and its correlation with \( {\rm Re} ( \epsilon' / \epsilon_K ) \), 
respectively. There is no definite correlation between 
$K_L \rightarrow \pi^0 \nu \bar{\nu}$ and
\( {\rm Re} ( \epsilon' / \epsilon_K ) \) in our model, unlike the scenario 
in which the new physics affects \( {\rm Re} ( \epsilon' / \epsilon_K ) \) 
by a modified $sdZ$ vertex. We note that one cannot make definite predictions 
for the branching ratio for this decay unless we know the angle $\gamma$ in 
the presence of new physics. The branching ratio may be vanishingly small or 
larger than the SM prediction by a factor of $\lesssim 2$. Therefore, this 
decay once measured may provide us with invaluable informations on the value 
of the angle $\gamma$, if there is no significant new physics contribution to
$s\rightarrow d \nu \bar{\nu}$ upto 4--fold ambiguities, which could be
eliminated using information from other rare kaon decays and $B$ meson decays.
In Fig.~\ref{fig:kpnn-mi5} (c) and (d), we show the similar plots
for $K^+ \rightarrow \pi^+ \nu \bar{\nu}$. Again, the measurement of the
$K^+ \rightarrow \pi^+ \nu \bar{\nu}$ branching ratio will help
to determine $\gamma$. And the branching ratio for this decay
may be larger than the SM predictions by a factor of $\sim 2$.


In Figs.~3 (a) and (b), we show the branching ratio of 
\( B( K_L \rightarrow \pi^0 e^+ e^- )_{\rm dir} \) 
as a function of $\phi$, 
the argument of $( \delta_{12}^d )_{LL}$, and its nontrivial correlation with 
\( {\rm Re} ( \epsilon' / \epsilon_K ) \). 
This nontrivial correlation arises because electrons couple to a photon
and also the \( s \rightarrow d \ga \) vertex contributes both
\( K_L \rightarrow \pi^0 e^+ e^- \) and 
\( {\rm Re} ( \epsilon' / \epsilon_K ) \).
In other words, the Wilson coefficient for $s\rightarrow d \gamma$
is modified by the gluino mediated FCNC, which is in turn closely related
with the gluino mediated $sdg$ operator. 
Also the Wilson
coefficient of $O_{7V}$
vertex is  modified  by the gluino mediated FCNC, which affects the 
direct CP violating part in $K_L \rightarrow \pi^0 e^+ e^- $.
This is in constrast with the $K\rightarrow \pi\nu\bar{\nu}$ case which
is not affected directly by gluino mediated FCNC.
Therefore, the direct CP violating part
of \( K_L \rightarrow \pi^0 e^+ e^- \) can be affected a lot,
if there is a new CP violating phase in $\tilde{s}_L -
\tilde{d}_L$ mixing both by a single MI of $(\delta_{12}^d )_{LL}$ and
a double MI of $( \delta_{12}^d )_{LR}^{\rm ind} = ( \delta_{12}^d )_{LL} 
\times ( \delta_{22}^d )_{LR}$. 
Also this process is affected indirectly 
by SUSY gluino effects 
through their dependence on the KM angle $\gamma$.  
After all, the branching ratio of $K_L \rightarrow \pi^0 e^+ e^-$ can
be vanishingly small or enhanced over the SM value by a factor of $\sim 3$.

Finally, we consider the gluino mediated SUSY contributions to the short
distance part of $ K_L \rightarrow \mu^+ \mu^-$. Again this decay
could be affected by gluino mediated FCNC both directly and indirectly :
through the modified Wilson coefficient of $sd\gamma$ operator and
modified values of $\gamma$ due to SUSY contributions to $\epsilon_K$.
However, unlike the direct CP violating part in
\( K_L \rightarrow \pi^0 e^+ e^- \), the modified $sd\gamma$ is irrelevant
to $K_L \rightarrow \mu^+ \mu^-$ because of current conservation associated
with the muon current. Therefore it depends only on the angle $\gamma$ 
[Fig.~4 (a)], and its correlation with \( \epsilon' / \epsilon \) is rather 
trivial [Fig.~4 (b)]. Depending on the values of $\gamma$, the result can be 
reduced by a factor of $\sim 3$ or increased by a factor of $\sim 2$.

Let us summarize the results on rare kaon decays within mass insertion
approximation. We assumed there are significant gluino--squark loop
contributions to $\epsilon_K$ and \( {\rm Re} ( \epsilon' / \epsilon_K ) \)
through $( \delta_{12}^d )_{LL} \sim O(10^{-2} - 10^{-3} )$ and
the $( \delta_{12}^d )_{LR} \sim O(10^{-5})$ induced by a double
mass insertion of flavor preserving $\tilde{s}_R - \tilde{s}_L$
followed by $( \delta_{12}^d )_{LL}$. Then these parameters
directly affect $s\rightarrow d g$ and $s\rightarrow d \gamma$
vertices, but not $s\rightarrow d \nu \bar{\nu}$. Also the
shape of the UT would be different from the SM case because
SUSY loop contributes to $\epsilon_K$. Therefore all the rare
kaon decays (and also $B$ decays as well, although we do not
discuss this subject in any detail here) are affected indrectly
through modified CKM elements, especially $\gamma$. For $K
\rightarrow \pi \nu\bar{\nu}$ and the short distance part of
$K_L \rightarrow \mu^+ \mu^-$, the branching ratios can be
larger than the SM expectations by a factor of $\sim 2-3$, or can be
smaller than the SM values. Especially the branching ratio for
$K_L \rightarrow \pi^0 \nu\bar{\nu}$ can be vanishingly small
if $\gamma = 0$ or $\pi$. Their correlations with
\( {\rm Re} ( \epsilon' / \epsilon_K ) \) are rather trivial, because 
these decays are not affected directly by $( \delta_{12}^d )_{LL} $.
On the other hand, the direct CP violating part of 
$K_L \rightarrow \pi^0 e^+ e^-$ is affected both directly and indirectly, 
and its correlation with \( {\rm Re} ( \epsilon' / \epsilon_K ) \)
is nontrivial at all. One may expect much enhanced (or reduced)
direct CP violations in $K_L \rightarrow \pi^0 e^+ e^-$ in our scenario,
depending whether SUSY effects can interfere constructively
(destructively) with the SM amplitude.



\section{Vertex Mixing Case}
\label{sec:vm}

\subsection{Flavor mixings in the quark-squark-gluino vertices}

Another way of solving the SUSY flavor and CP problem is to assume that 
the 1st/2nd generation squarks are heavy and there are certain degrees of 
hierarchy in the $q_{i}-\tilde{q}_{j}-\tilde{g}$ mixing matrices, $W_L$ and 
$W_R$ (the decoupling scenario).  Here $W_L$ and $W_R$ are analogous to the 
CKM matrix in the SM. Interactions among quarks, squarks, and gluinos in 
terms of their mass eigenstates are described by the Lagrangian:\footnote{%
Our definition of the mixing matrix
$W$ is different from that in Ref.~\cite{barbieri:97} in that the
signs of the two terms are opposite and both $W_L$ and $W_R$ are
not complex-conjugated.}
\begin{equation}
     {\cal L} = - \sqrt{2} \, g_s (W_L)_{ij} \,
                  \wt{q}_L^{*i} \, \ol{\wt{g}}_R^a t^a q_L^j
                + \sqrt{2} \, g_s (W_R)_{ij} \,
                 \wt{q}_R^{*i} \, \ol{\wt{g}}_L^a t^a q_R^j
                + {\rm h.c.},
\end{equation}
where $i$ and $j$ are generation indices and color indices of
(s)quarks have been omitted.  
Although squarks of each chirality are diagonalized into their
(approximate) mass eigenstates, the remaining small mixing between 
squarks of different chiralities can be treated as a perturbation
except for the stop sector.

\subsection{Wilson coefficient for $\Delta S =2$}

The Wilson coefficients for $\Delta S = 2$ effective Hamiltonian in the
VM case are given by following expressions : 
   \beqar             
   C^{(2)}_1&=&+\frac{1}{4} \left( 24  f_4(x)+ 66 \wt{f}_4(x) \right)
         \left( F_{12} \right)^2_{LL}, \nonumber \\
  C^{(2)}_2&=&+\frac{204}{4}  f_4(x) 
   \left( F_{12} \right)^2_{RL}, \nonumber \\
   C^{(2)}_3&=&-\frac{36}{4}  f_4(x) 
   \left( F_{12} \right)^2_{RL}, \nonumber \\
   C^{(2)}_4&=&+\frac{1}{4} \left( 504  \wt{f}_4(x)- 72 f_4(x) \right)
         \left( F_{12} \right)_{LL} 
         \left( F_{12} \right)_{RR} -
         \frac{132}{4} \wt{f}_4(x) \left( F_{12} \right)_{LR}
                  \left( F_{12} \right)_{RL},  \nonumber \\
   C^{(2)}_5&=&+\frac{1}{4} \left( 24  \wt{f}_4(x)+ 120 f_4(x) \right)
         \left( F_{12} \right)_{LL} 
         \left( F_{12} \right)_{RR} -
         \frac{180}{4} \wt{f}_4(x) \left( F_{12} \right)_{LR}
            \left( F_{12} \right)_{RL},  \nonumber \\
   \wt{C}^{(2)}_1&=&+\frac{1}{4} \left( 24  f_4(x)+ 66 \wt{f}_4(x) \right)
         \left( F_{12} \right)^2_{RR}, \nonumber \\
   \wt{C}^{(2)}_2&=&+\frac{204}{4}  f_4(x) 
   \left( F_{12} \right)^2_{LR}, \nonumber \\
   \wt{C}^{(2)}_3&=&-\frac{36}{4}  f_4(x) \left( F_{12} \right)^2_{LR},
   \eeqar         
   where
   \beqar             
   f_4(x)&=& x~\frac{2x-2-(1+x)\ln x}{(x-1)^3} ,
   \nonumber \\
   \wt{f}_4(x)&=&\frac{x^2-1-2x\ln x}{(x-1)^3} ,
   \nonumber \\
   \left( F_{12} \right)_{AB} &=& (W_{A})^{*}_{31} (W_{B})_{32}.~~~~~~~~
   ( A, B = L~ {\rm or}~ R )
   \eeqar
   Here we assume that both left and right sbottom masses are equal
   to $\tilde{m}$,
   and define $ x = ( m_{\tilde{g}}/\tilde{m} )^2$.
%

\subsection{Wilson coefficient for $\Delta S =1$}

The Wilson coefficients for $\Delta S = 1$ effective Hamiltonian in the
VM case are given by following expressions :       
   \beqar
   C_i &=& 0~~~~~~ \mbox{(for $i$ = 1,2,7 $\sim$ 10)}, \nonumber \\
   C_3 &=& \frac{1}{4} ~\frac{\alpha_s^2}{2\wt{m}^2}
          (F_{12})_{LL} P(x), \nonumber \\
   C_4 &=& - \frac{1}{4} ~\frac{3\alpha_s^2}{2\wt{m}^2}
          (F_{12})_{LL} P(x), \nonumber \\
   C_5 &=& \frac{1}{4} ~\frac{\alpha_s^2}{2\wt{m}^2}
          (F_{12})_{LL} P(x), \nonumber \\
   C_6 &=&-  \frac{1}{4} ~\frac{3\alpha_s^2}{2\wt{m}^2}
          (F_{12})_{LL} P(x), \nonumber \\
   \wt{C}_i&=& C_i ~\mbox{with replacement of $LL$ by $RR$},
   \quad i = 3, 4, 5, 6 \nonumber \\
   C_{\gamma s}^\pm
   &=& \frac{8 \pi Q_d}{3\wt{m}^2} 
   \left[ (F_{12})_{LL} \pm
     (F_{12})_{RR}  \right] P_{BE}(x), \\
   C_{\gamma \tilde{g}}^\pm 
   &=& \frac{8 \pi Q_d}{3\wt{m}^2}
   \left[
     (F_{12})_{LR} \frac{m_b (A^*_b -\mu \tan \beta)}{\wt{m}} 
      \pm
       (F_{12})_{RL} \frac{m_b (A_b -\mu^* \tan \beta)}{\wt{m}} 
            \right] \sqrt{x} P_{BI}(x,x), 
   \\
   C_{g s}^\pm 
   &=& \frac{2 \pi}{\wt{m}^2} \left[ (F_{12})_{LL}  \pm
     (F_{12})_{RR}  \right]  P_2(x), \\
   C_{g \tilde{g}}^\pm
   & = &
   \frac{2 \pi}{\wt{m}^2}
   \left[
     (F_{12})_{LR} \frac{m_b (A^*_b -\mu \tan \beta)}{\wt{m}} 
     \pm
     (F_{12})_{RL} 
     \frac{m_b (A_b -\mu^* \tan \beta)}{\wt{m}}
     \right] \sqrt{x} P_3(x,x),
   \eeqar
where the $LR$ mixing parameter $( A_b^* - \mu \tan \beta )$ is real, and
   \beqar
   P_F(x) &=& \frac{1}{36} \frac{1}{(1-x)^4} \left[ 7 x^3 -36 x^2 +45 x -16
    + (18 x -12) \ln x \right], \nonumber\\
   P_B(x) &=& \frac{-1}{36} \frac{1}{(1-x)^4}  \left[ 11 x^3 -18 x^2 +9 x -2
   - 6 x^3 \ln x \right], \nonumber \\
   P_{FE}(x) &=& \frac{1}{12} \frac{-1}{(1-x)^4} \left[  x^3-6x^2+3x+2
   + 6x \ln x \right], \nonumber \\
   P_{BE}(x) &=& \frac{1}{12} \frac{1}{(1-x)^4} \left[ 2 x^3+3x^2-6x+1-
    6x^2 \ln x \right], \nonumber \\
   P_{FI}(x) &=& \frac{1}{2} \frac{- x}{(1-x)^3} \left[ 4 x -x^2-3- 2
   \ln x \right], \nonumber \\
   P_{BI}(x) &=& \frac{1}{2} \frac{x}{(1-x)^3} \left[ 1-x^2+2x \ln x 
   \right], \nonumber \\
   P(x)&=& P_F(x) - \frac{1}{9} P_B(x), \nonumber \\
   P_2(x) &=& \frac{3}{2}P_{FE}(x) -\frac{1}{6}P_{BE}(x), \nonumber \\
   P_3(x) &=& \frac{3}{2}P_{FI}(x) -\frac{1}{6}P_{BI}(x), \nonumber \\
   P_{BI}(x,x)&=&\lim_{y\to x}\frac{P_{BI}(x)-P_{BI}(y)}{x-y}, \nonumber\\
   P_3(x,x)&=&\lim_{y\to x}\frac{P_3(x)-P_3(y)}{x-y}. 
   \eeqar
   We don't consider the box diagrams in vertex mixing, because it will be 
   highly suppressed due to heavy 1st/2nd generation squarks.  

   The Wilson coefficient of $O^{\rm SUSY \prime}_{7V}$ is
   \begin{equation}
     C^{\rm SUSY  \prime}_{7V} =
     - \frac{4}{9 \tilde{m}^2} P_B (x) (F_{12})_{LL} .
   \end{equation}

\subsection{EDM constraint}

   The Wilson coefficients for the effective Hamiltonian for the
   neutron EDM in VM are given by
   \begin{eqnarray}
     C_1^{\rm edm} & = &
     - \frac{2}{3} \: \frac{e \al_s}{\pi} \: Q_d \:
     \frac{\mg}{\wt{m}^2} \:
     \frac{m_b \: {\rm Im} \left[ (F_{11})_{LR} (A_b^* - \mu \tan \be) \right]}
     {\wt{m}^2} \:
     P_{BI} (x, x) , \nonumber \\
     C_2^{\rm edm} & = &
     - \frac{1}{4} \: \frac{g_s \al_s}{\pi} \:
     \frac{\mg}{\wt{m}^2} \:
     \frac{m_b \: {\rm Im} \left[ (F_{11})_{LR} (A_b^* - \mu \tan \be) \right]}
     {\wt{m}^2} \:
     2 \left(\frac{3}{2} P_{FI} - \frac{1}{6} P_{BI} \right) (x, x) .
   \end{eqnarray}
   Basically, the constraint on \( {\rm Im} (\delta_{11}^d)_{LR} \)
   in (\ref{eq:Imd11LR}),
   applies to
   \( 
   m_b \: {\rm Im} \left[ (F_{11})_{LR} (A_b^* - \mu \tan \be) \right] /
   \wt{m}^2
   \)
   much the same way,
   and if we assume \( (A_b^* - \mu \tan \be) \) is real and is about
   $2 \: {\rm TeV}$, this limits
   \begin{equation}
     \label{eq:Imblah}
     {\rm Im} (F_{11})_{LR} \: \lesssimilar \:
     {\cal O} (10^{-7}) .
   \end{equation}
   Since we will set $(W_R)_{31}$ to zero
   in the numerical analysis, there is no difficulty
   in satisfying this limit in our case, but it should be obeyed
   when other possible solutions are searched for.
   For example, if both $(W_L)_{31}$ and $(W_R)_{31}$
   have absolute values of
   ${\cal O} (10^{-2})$, the typical order of magnitude of $| (W_L)_{31} |$
   in our analysis, then their phase difference should be
   less than ${\cal O} (10^{-3})$.


\subsection{Numerical Results}

In MI case, the only key parameter responsible for CP violation was
\( (\delta_{12}^d)_{LL} \). Once its modulus is determined by $\epsilon_K$ 
constraint for a fixed $\gamma$, every CP violating observable  could be 
computed as a function of its phase, ${\rm Arg} (\delta_{12}^d)_{LL}$. 
Within the VM approach, on the contrary, the analysis becomes more 
complicated, because we have four independent complex parameters,
$(W_L)_{31}$, $(W_L)_{32}$, $(W_R)_{31}$, and $(W_R)_{32}$, that describe 
the flavor mixing in the $q-\tilde{q}-\tilde{g}$ with various chiralities 
of quarks and their superpartners.  Even if we impose the \( \epsilon_K \) 
and the EDM constraints, there still remain 6 real parameters and it would 
be a formidable task to scan all of this 6 dimensional parameter space.
In order to simplify the numerical analysis, we discarded some parameters 
and tried to find a particular solution rather than look for a completely 
general solution. That would not much distort the picture of VM approach 
and would suffice to illustrate a possible existence of appropriate parameter 
space in which gluino--squark loop contributions to $\epsilon_K$ and 
\( {\rm Re} (\epsilon'/\epsilon_K ) \) 
are significant without conflict with electron/neutron EDM's.

For definiteness, we set $(W_L)_{32}$ and $(W_R)_{31}$ to zero, and all 
SUSY contributions are expressed in terms of a single complex number 
$(F_{12})_{LR} \equiv ( W_L )_{31}^* ( W_R )_{32}$. The results for 
\( {\rm Re} ( \epsilon' / \epsilon_K ) \) for various values of the angle 
$\gamma$'s are displayed in Figs.~\ref{fig:epsp-phase-vm2}, as functions 
of $\phi$, where we use the following parameterization,
\( (F_{12})_{LR} = r e^{i \phi} \). As in the MIA case, we can have a large 
\( \epsilon' / \epsilon \) in certain region of $\phi$ with size 
$\sim {\cal O}(1)$. 
In particular, all the CP violations in the kaon system could 
be accommodated in terms of CP violations from SUSY sector with vanishing 
KM angle, $\gamma = 0$ or $\pi$, 
if $A_b^* - \mu \tan\beta = {\cal O}(2) $ TeV.  
Therefore, $\tan\beta$ could be considerably smaller than the MIA case 
because
the flavor preserving $LR$ mixing in the VM case is enhanced by 
$m_b /m_s$ relative to that in the MIA.
Also SUSY particle mass spectra will 
be quite different from those in the MIA case, although both cases can 
accommodate the CP violations in the kaon system. 


The decays $K\rightarrow \pi \nu \bar{\nu}$ and the short distance part
of $K_L \rightarrow \mu^+ \mu^-$ in the VM will be  essentially the same 
as those in the MIA. There are no direct SUSY effects on these processes.
SUSY effects are only indirect through the angle $\gamma$ from the SUSY 
contributions to $\epsilon_K$.  Therefore we do not show any plots for
these processes. The $\gamma$ dependence are the same as the MIA
[see Figs.~2 (a), (c) and Fg.~4 (a)].

\begin{figure}
  \begin{center}
    \subfigure[]{
      \includegraphics[width=7.5cm]{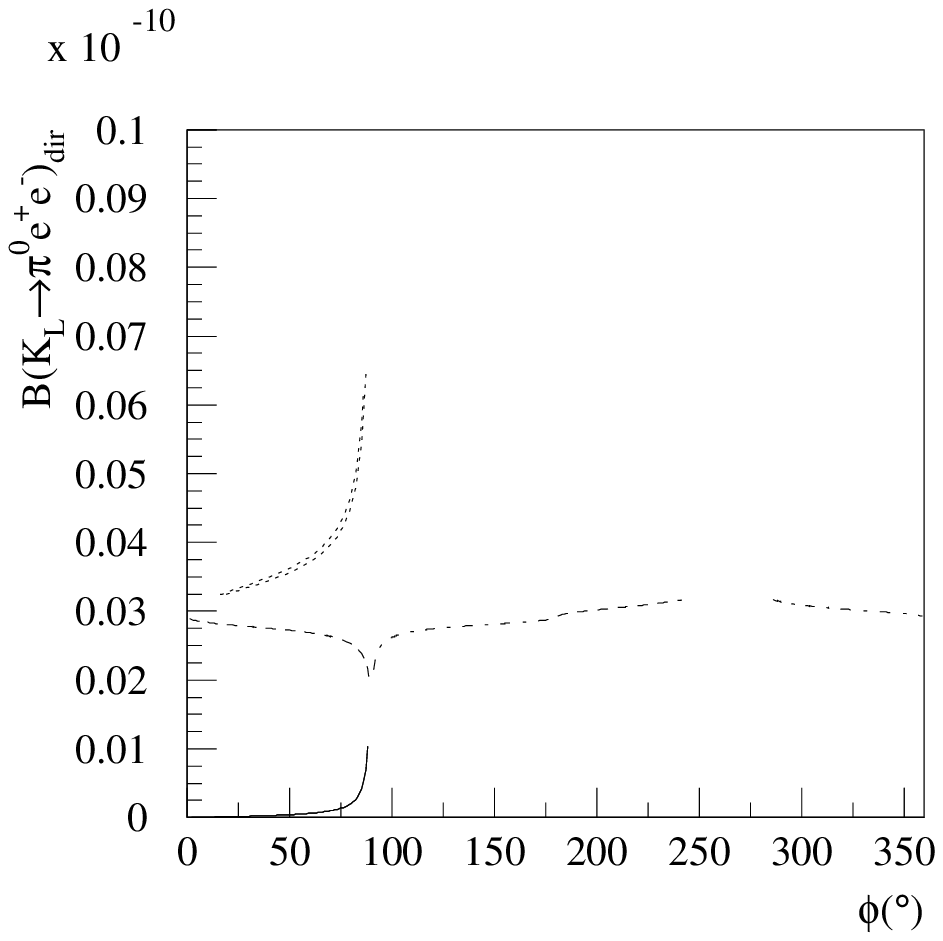}}
    \subfigure[]%
    {\includegraphics[width=7.5cm]{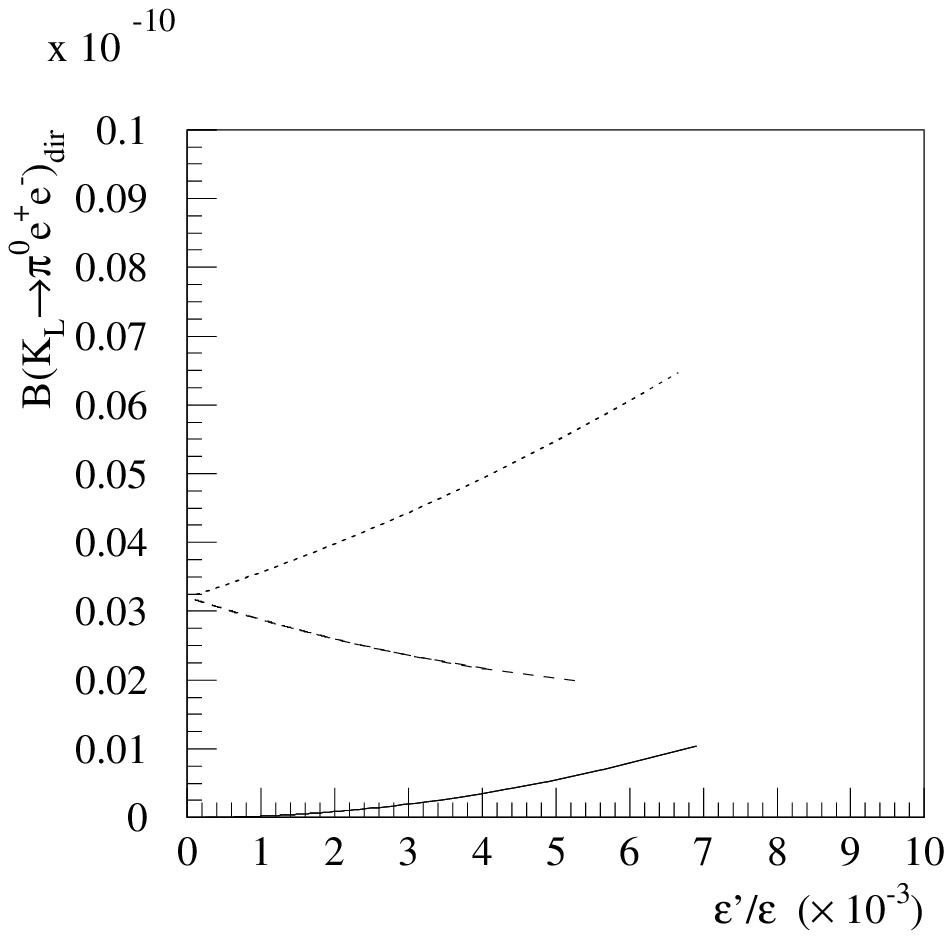}}
    \caption{
      The $\phi$ dependence of
      $ B( K_L \rightarrow \pi^0 e^+ e^- )_{\rm dir} $ ((a))
      and its correlation with
      $ \epsilon' / \epsilon $ ((b)), in MI for $x = 1$ with
      $A_s^* - \mu \tan \be = 10 \: {\rm TeV} $.
      (a)
      The solid, the dashed, the dotted, and the dash-dotted lines
      correspond to
      $ \ga = 0^\circ, 180^\circ; 60^\circ; 240^\circ, 300^\circ;
      120^\circ $; respectively.
      (b)
      The solid, the dashed, and the dotted lines correspond to
      $ \ga = 0^\circ, 60^\circ, 240^\circ$,
      respectively, and $180^\circ$ minus them.
      }
    \label{fig:kpee-mi10}
  \end{center}
\end{figure}

On the contrary, the direct CP violating part of 
$K_L \rightarrow \pi^0 e^+ e^-$ is affected directly by the SUSY loop 
contributions. In Fig.~\ref{fig:kpee-vm2}~(a), 
we show its branching ratio as a function
of the angle $\phi$ for different values of $\gamma$. 
In Fig.~\ref{fig:kpee-vm2}~(b), we 
show its correlation with \( {\rm Re} (\epsilon' / \epsilon_K )\). 
It
looks very similar to Fig.~\ref{fig:kpee-mi10}~(b) because
\( C_\ga^+ = C_\ga^- \) and \( C_g^+ = C_g^- \) here
as was in the MIA case.
If we relax the assumption that $(W_L)_{32} = (W_R)_{31} = 0$, however,
then
the correlation may differ from that in the MIA case in general because
\( {\rm Re} (\epsilon' / \epsilon_K )\) mainly depends on
$C_g^-$ while \( B(K_L \rightarrow \pi^0 e^+ e^-) \) on $C_\ga^+$.

\section{ Conclusions }
\label{sec:conc}

In this paper, we discussed the gluino--squark loop contributions to CP 
violations in the kaon system both in the MIA and in the VM. 
Unlike many other models based on the enhanced $sdg$ or $sdZ$ vertex in 
the context of SUSY using nonuniversal flavor structures in the trilinear 
$A$ couplings, our model modifies both the $\eps$ and Re ($\epsp$), so 
that the phenomenological implications on other kaon decays substantially  
differ from other models. In particular the shape of unitarity triangle 
can be very different from the SM case, due to potentially large SUSY 
contributions to $\epsilon_K$ and $B^0 - \overline{B^0}$ mixing. Only the 
constraint from the semileptonic $b\rightarrow u$ transition is stable 
against a possible SUSY contributions. Therefore we varied the $\gamma$ 
from $0$ to $2 \pi$, then fixed the $\lambda_t$ to get the right magnitude 
of $\eps_K$.  In particular, one can accommodate both $\epsilon_K$ and 
\( {\rm Re} ( \epsilon^{'} / \epsilon_K ) \), even if the angle $\gamma$ 
is very close to $0$ or $\pi$, still consistent with the ranges obtained 
from the three generation unitairity relations, Eq.~(2).
In our model, the gluino--squark loop does modify $s\rightarrow d \gamma$ 
just like  the $sdg$ operator, but does not directly contribute to 
$s\rightarrow d \nu \bar{\nu}$.  Still the branching ratios for 
$K \rightarrow \pi \nu \bar{\nu}$ can be different from the SM predictions 
by a large amount because of the modified $\gamma$. Similar effects occur in 
$K_L \rightarrow \pi^0 e^+ e^-$ and $K_L \rightarrow \mu^+ \mu^-$ processes.  
If the shape of the triangle in MSSM is the same as in the SM case, then the 
deviations from the SM predictions are possible only in $\epsp$ and 
$K_L \rightarrow \pi^0 e^+ e^-$, although the deviation in the latter case
is very small. Also in our model, SUSY effects give additional contributions 
to $\Delta S = 1$ penguin operators, so that we may anticipate interesting 
deviations of direct CP asymmetries in $K \rightarrow 3 \pi$ and 
$K \rightarrow \pi \pi \gamma$ from the SM predictions. However, these 
observables are generically contaminated by hadronic uncertainties even in 
the chiral perturbation theory, and are left for the future study.  

One can also perform  a similar analysis in the $B$ meson sector
including SUSY gluino--squark loop contributions to 
$B^0 - \overline{B^0}$ mixing, and considering a specific flavor structure
for the mass insertion parameters \cite{work}.  Depending on the chiral 
structures of the mass insertion parameters, one could achieve 
$\sin 2\beta_{J/\psi K_s}$ which is very different from the SM case.
In this case, the results could be very different from the SM even if we 
assume that the UT in the MSSM is the same as in the SM. 

Our study indicates that the measurements of 
$K_L \rightarrow \pi^0 \nu \bar{\nu}$ in the kaon sector is very important 
for constructing the unitarity  triangle, especially if there are large new 
physics contributions to $\epsilon_K$ and $B^0 - \overline{B^0}$ mixing. 
Also, in the $B$ meson sector, it would be utterly important to construct 
the UT from the tree level processes which are immune to possible new 
physics contributions. For example, extracting $\ga$ from $B \rightarrow D K$ 
or other tree dominated decays will be very desirable \cite{bdk}.  
The current global analysis on CKM matrix elements suggests that a unique
UT emerges and the SM picture for flavor and CP violations in the quark sector
seems to be a correct picture. But as we demonstrated with our specific model,
this is no longer guaranteed in the presence of new physics which give 
significant contributions to $\epsilon_K$ and $B^0 - \overline{B^0}$
mixing \cite{kn2}.

\acknowledgements

This work is supported in part by BK21 project of the Ministry of Education
and SRC of KOSEF (PK, JP) and by Chonbuk National University(JJ).


\begin{figure}
  \begin{center}
    \subfigure[]{
      \includegraphics[width=7.5cm]{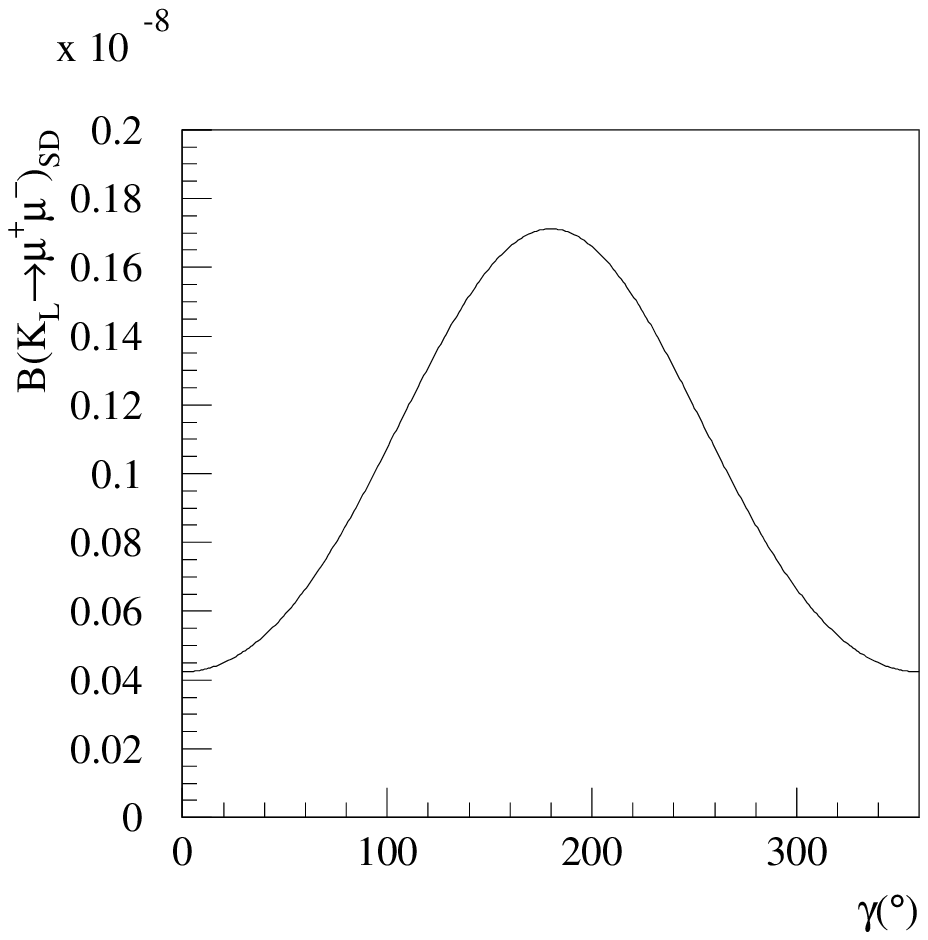}}
    \subfigure[]{
      \includegraphics[width=7.5cm]{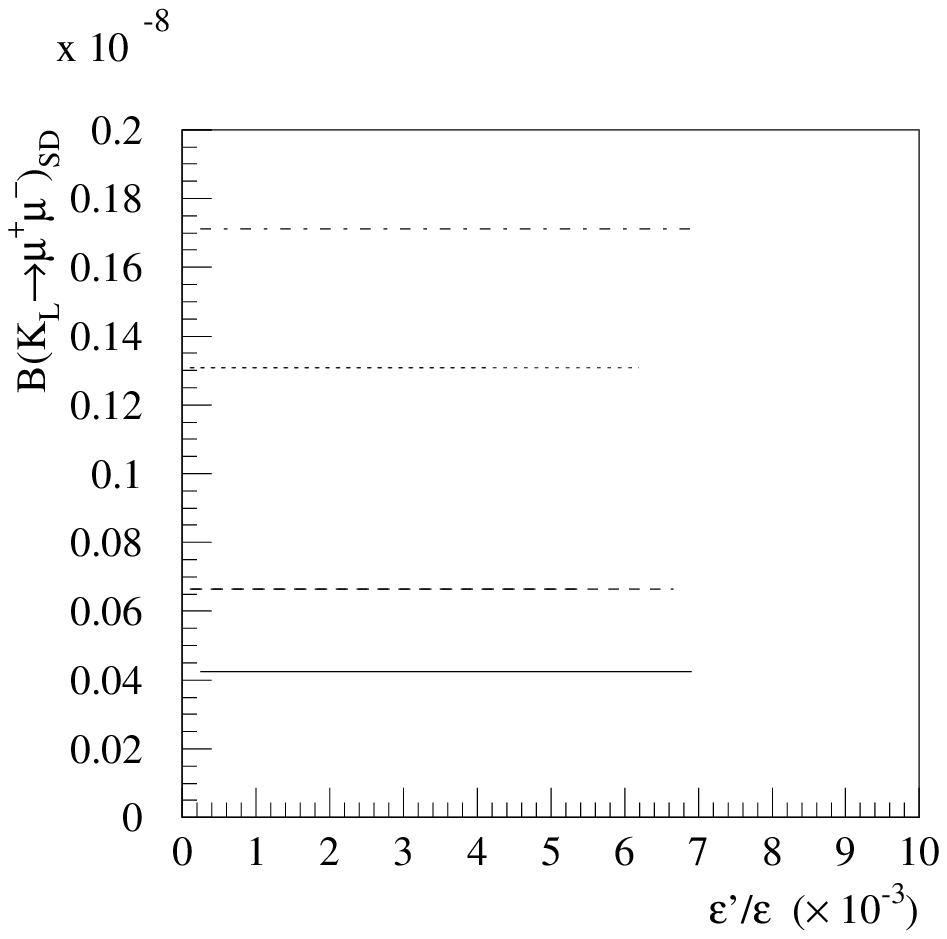}}
    \caption{
      The $\gamma$ dependence of
      $ B( K_L \rightarrow \mu^+ \mu^- )_{\mtext{SD}} $ ((a))
      and its correlation with
      $ \epsilon' / \epsilon $ in MI ((b)), for $x = 1$ with
      $A_s^* - \mu \tan \be = 10 \: {\rm TeV} $.
      (b)
      The solid, the dashed, the dotted, and the dash-dotted lines
      correspond to
      $ \ga = 0^\circ, 60^\circ, 240^\circ, 180^\circ $,
      respectively, and $360^\circ$ minus them.
      }
    \label{fig:kmm-mi5}
  \end{center}
\end{figure}

\begin{figure}
  \begin{center}
      \includegraphics[width=17cm]{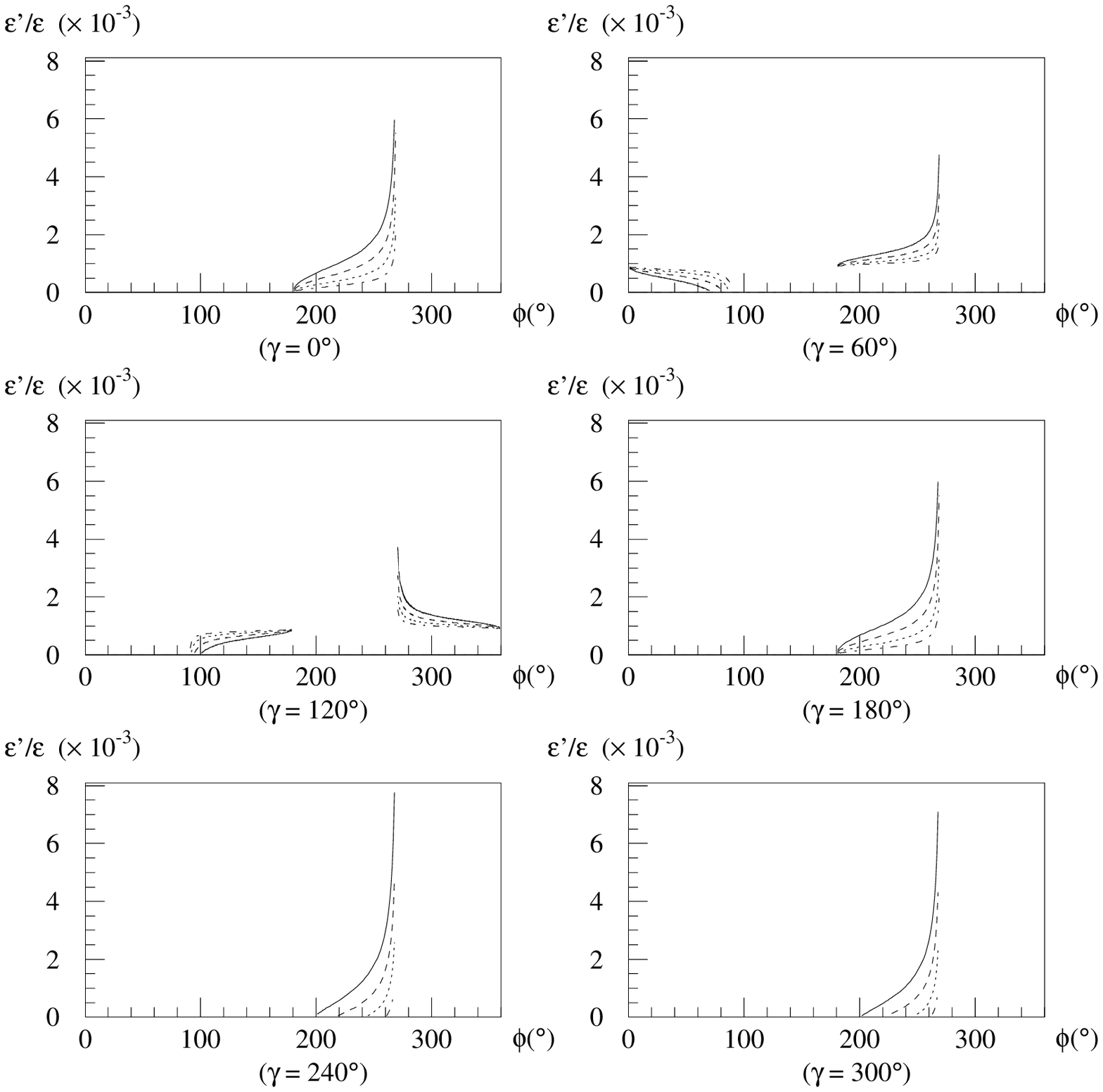}
    \caption{
      The plots of
      $ \epsilon' / \epsilon $ versus the phase of
      $(F_{12})_{LR}$ in VM for
      six different values of $\ga$ with
      $A_b^* - \mu \tan \be = 2 \: {\rm TeV} $.
      Graphs were drawn in the solid, the dashed, the dotted,
      and the dash-dotted lines
      for $ x = 0.5, 1.0, 2.0, 4.0 $, respectively.
      }
    \label{fig:epsp-phase-vm2}
  \end{center}
\end{figure}


\begin{figure}
  \begin{center}
    \subfigure[]{
      \includegraphics[width=8cm]{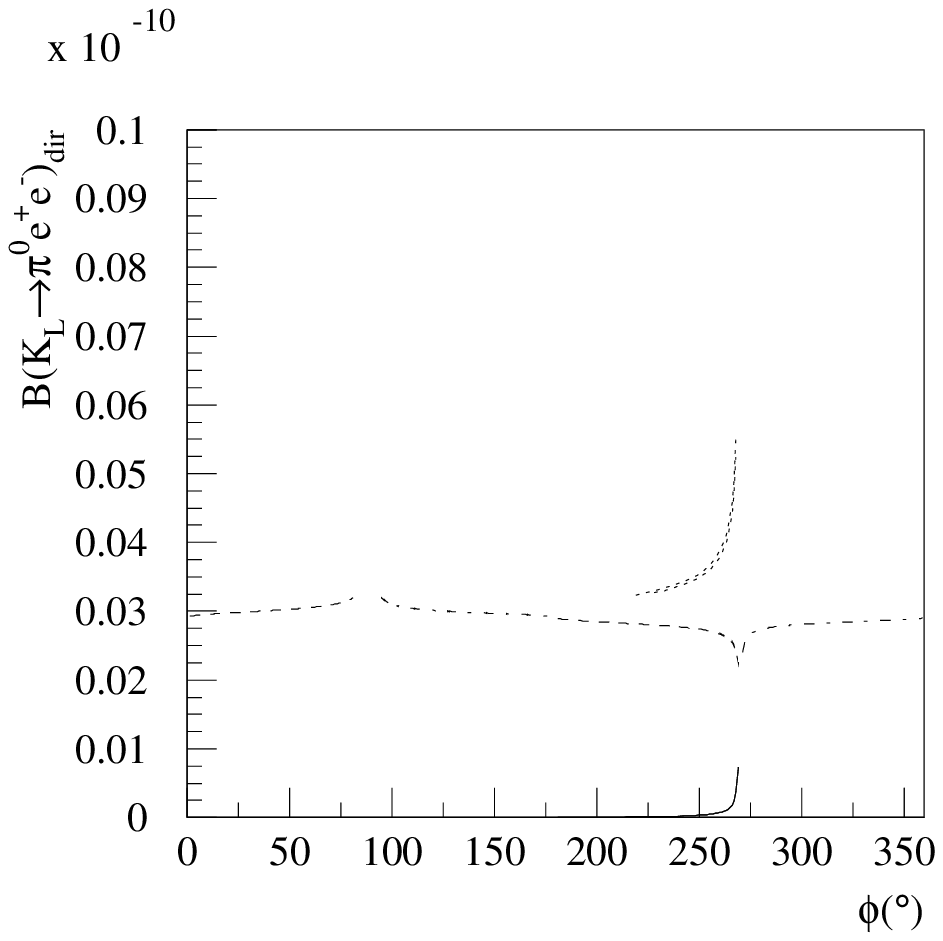}}
    \subfigure[]{
      \includegraphics[width=8cm]{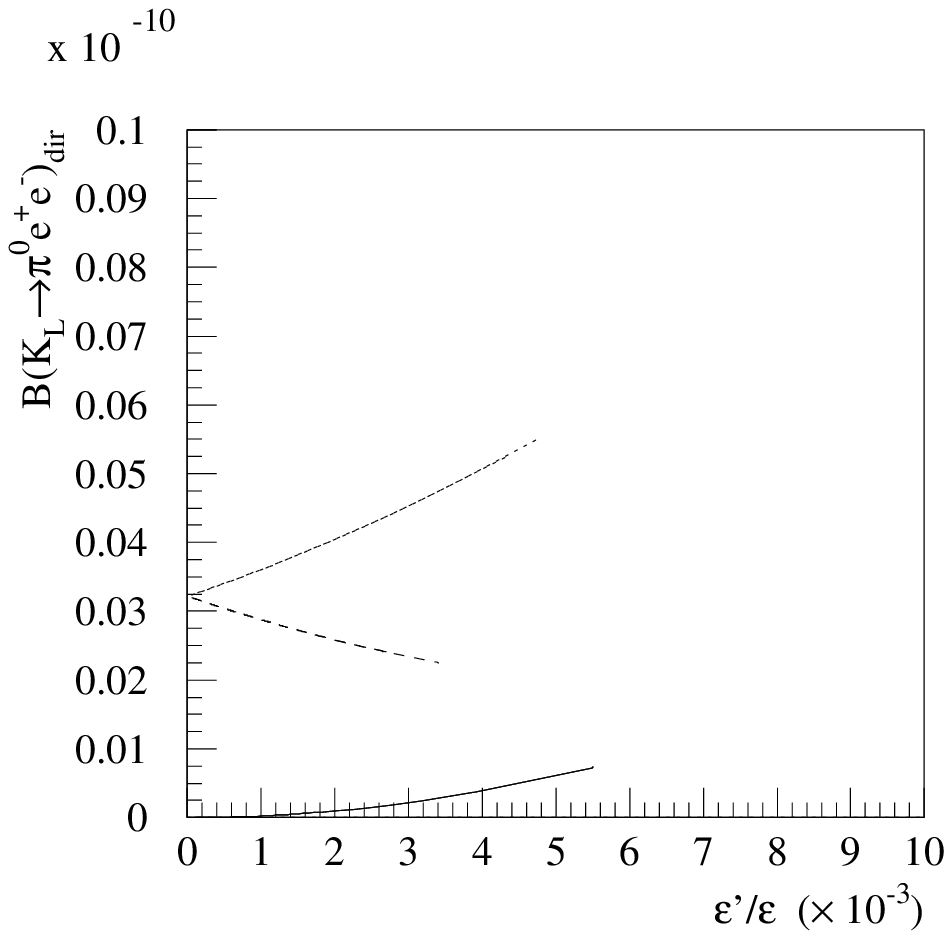}}
    \caption{
      The $\phi$ dependence of
      $ B( K_L \rightarrow \pi^0 e^+ e^- )_{\rm dir} $ ((a))
      and its correlation with
      $ \epsilon' / \epsilon $ ((b)), in VM for $x = 1$ with
      $A_s^* - \mu \tan \be = 2 \: {\rm TeV} $.
      These curves are specific to the solution of parameter space
      drawn in Fig.~\ref{fig:epsp-phase-vm2} which restricts
      $(W_L)_{32}$ and $(W_R)_{31}$ to zero.
      The meanings of line patterns are
      the same as in Fig.~\ref{fig:kpee-mi10}.
      }
    \label{fig:kpee-vm2}
  \end{center}
\end{figure}

\begin{table}
  \begin{center}
    \begin{minipage}{13cm}
    \begin{tabular}{c|r||c|l}
      \hspace{5mm}$m_Z$\hspace{5mm} & 91.2 GeV\hspace{5mm} & $\alpha_s(m_Z)$\hspace{5mm} & 0.118 \phantom{MeV}\hspace{5mm} \\              
      $m_W$ & 80.2 GeV\hspace{5mm} & $f_{\pi}$\hspace{5mm} & 131 MeV\hspace{5mm} \\                                
      $m_t$ & 170  GeV\hspace{5mm} & $f_K$\hspace{5mm} & 160 MeV\hspace{5mm} \\                                    
      $m_b$ & 4.4  GeV\hspace{5mm} & $\Delta M_K$\hspace{5mm} & $3.51 \times 10^{-15}$ GeV\hspace{5mm} \\          
      $m_c$ & 1.3  GeV\hspace{5mm} & $\sin^2 \theta_W$\hspace{5mm} & 0.23 \phantom{MeV}\hspace{5mm} \\             
      $m_d$ & 8    MeV\hspace{5mm} & $|\epsilon_K|$\hspace{5mm} & $(2.266 \pm 0.023) \times 10^{-3}$ \phantom{MeV}\hspace{5mm} \\
      $m_\pi$ &135 MeV\hspace{5mm} & $({\rm Re} A_0)_{\rm exp}$\hspace{5mm} & $3.33 \times 10^{-7}$ \phantom{MeV\hspace{5mm}} \\
      $m_K$ & 498  MeV\hspace{5mm} & $({\rm Re} A_2)_{\rm exp}$\hspace{5mm} & $1.50 \times 10^{-8}$ \phantom{MeV}\hspace{5mm}
    \end{tabular}
    \end{minipage}
    \vspace{.5cm}
  \caption{Input values we used in the numerical analysis.}
  \label{tab:input}
  \end{center}
\end{table}


\end{document}